\documentclass[english,11pt]{article}

\pdfoutput=1

\usepackage{amsmath}
\usepackage{amsfonts}
\usepackage{amssymb}
\usepackage{amstext} 
\usepackage{xfrac}
\usepackage{graphicx, rotating}
\usepackage{epstopdf}
\usepackage{epsfig}
\usepackage{latexsym}
\usepackage{multirow}
\usepackage[dvipsnames]{xcolor}
\usepackage{slashed}
\usepackage[top=2.8 cm, bottom=2.8 cm, left=2.5 cm, right=2.5 cm]{geometry}
\usepackage{array}  
\usepackage{hyperref}
\hypersetup{
	colorlinks = true,
	linkcolor = {Salmon},
	citecolor= {SeaGreen},
	urlcolor= {MidnightBlue}
}
\usepackage{braket}
\usepackage{dsfont}
\usepackage[compat=1.0.0]{tikz-feynman}
\usepackage{comment}
\usepackage{scalerel}
\usepackage{enumerate}
\usepackage{appendix}
\usepackage{tabularx}
\usepackage{booktabs}
\usepackage{multirow}
\usepackage{caption}
\usepackage{subcaption}
\usepackage{bm}
\usepackage{lipsum}
\usepackage{soul}

\usepackage[nottoc]{tocbibind}
\usepackage{cite}

\numberwithin{equation}{section}

\renewcommand{\arg}{\text{arg}}

\newcommand{\daga}{^{\scalerel*{\dagger}{X}}}

\newcommand{\tr}{\text{tr}}

\newcommand{\gev}{\text{ GeV}}

\newcommand{\lagr}{\mathcal{L}}

\newcommand\blfootnote[1]{%
	\bgroup
	\renewcommand\thefootnote{\fnsymbol{footnote}}%
	\renewcommand\thempfootnote{\fnsymbol{mpfootnote}}%
	\footnotetext[0]{#1}%
	\egroup
}
\newcommand{\no}{\nonumber}

\newcommand{\ba}{\begin{eqnarray}}
	\newcommand{\ea}{\end{eqnarray}}

\begin{document}
\unitlength = 1mm

\thispagestyle{empty} 
\begin{center}
\vskip 3.4cm\par
{\par\centering \textbf{\LARGE Grand Color Axion}}

\vskip 1.2cm\par
{\scalebox{.85}{\par\centering \large  
\sc\hyperlink{AV}{\color{black}Alessandro Valenti}$\,^{a,b}$, \hyperlink{LV}{\color{black}Luca Vecchi}$\,^{b}$, \hyperlink{LXX}{\color{black}Ling-Xiao Xu}$\,^{a,b}$}
{\par\centering \vskip 0.7 cm\par}
{\sl 
$^a$~Dipartamento di Fisica e Astronomia ``G.~Galilei", Università di Padova, Italy
}\\
{\par\centering \vskip 0.2 cm\par}
{\sl 
$^b$~Istituto Nazionale di Fisica Nucleare, Sezione di Padova, I-35131 Padova, Italy
}\\

{\vskip 1.65cm\par}}
\end{center}

\begin{abstract}

We present a model that solves the strong CP problem via an axion parametrically heavier than the standard one. Within this picture the Standard Model quarks are embedded into a larger non-abelian Grand Color group that at high scales splits into ordinary QCD and an additional confining dynamics under which exotic chiral fermions are charged. Crucially, the vacuum expectation value of the axion is automatically relaxed to zero because the only renormalizable source of explicit CP violation, beyond those encoded in the topological angles, is contained in the Standard Model Yukawa couplings, and is therefore very suppressed. The Grand Color axion potential is controlled by the scale of the new confining group and is much larger than the QCD contribution, such that its dynamics is less exposed to the so-called ``axion quality problem". Potentially observable corrections to the effective topological angle can also arise, in our model as well as in a large class of heavy axion scenarios, from non-renormalizable Peccei-Quinn-{\emph{conserving}} interactions, which introduce a new ``heavy axion quality problem". Our model has a very minimal field content, it relies entirely on gauge invariance and does not require the introduction of additional symmetries beyond the usual one postulated by Peccei and Quinn. The phenomenology is very rich and can be tested at colliders as well as via cosmological observations. A particularly interesting portion of parameter space predicts a visible Grand Color axion of mass above the GeV and decay constant larger than a few TeV.

\end{abstract}

\blfootnote{
\hypertarget{AV}{\href{mailto:alessandro.valenti@pd.infn.it}{\color{black}{alessandro.valenti@pd.infn.it}}}
}
\blfootnote{
	\hypertarget{LV}{\href{mailto:luca.vecchi@pd.infn.it}{\color{black}{luca.vecchi@pd.infn.it}}}
}
\blfootnote{
	\hypertarget{LXX}{\href{mailto:lingxiao.xu@unipd.it}{\color{black}{lingxiao.xu@unipd.it}}}
}

\newpage
\setcounter{page}{1}
\setcounter{footnote}{0}
\setcounter{equation}{0}
\noindent

{
	\hypersetup{linkcolor=black}
	\tableofcontents
}


\section{Introduction}
\label{sec:introduction}

The QCD axion~\cite{Peccei:1977hh,Weinberg:1977ma,Wilczek:1977pj} is by far the most popular solution of the strong CP problem. There are at least three good reasons why this is the case:
\begin{itemize}
\item[(i)] It has a very simple setup;
\item[(ii)] It allows arbitrary sources of CP violation suppressed by scales much larger than 1 GeV;
\item[(iii)] It can be tested via a wide array of probes.
\end{itemize}
The QCD axion just requires the existence of an approximate global $U(1)_{\rm PQ}$, anomalous under QCD, which gets spontaneously broken at some high scale $f_a$. The minimality of these assumptions is envied by all alternative solutions to the strong CP problem, which instead demand more complicated structures in the UV and are therefore viewed as less plausible.

The second feature stems from the fact that the solution delivered by the QCD axion is active at scales of order $\Lambda_{\rm QCD}\sim1$ GeV. Whatever fundamental source of CP violation is present at short distances, when we run down to $\sim\Lambda_{\rm QCD}$ it gets completely encoded in the QCD topological angle $\bar\theta$ and higher-dimensional operators. By relaxing $\bar\theta$ to zero, the QCD axion completely removes the largest source of CP-violation within the low energy effective field theory.~\footnote{The sources of CP-violation contained in higher-dimensional operators suppressed by the weak scale and the new physics scale $\Lambda_{\rm UV}$, are easily kept under control. Operators suppressed by the weak scale lead to very small effects because of the peculiarity of the SM Yukawa sector. The operators suppressed by $\Lambda_{\rm UV}$ can be made parametrically smaller by decoupling the new physics from the QCD scale. Below we will discuss in detail the analogous effects that arise in our model.} This is a truly remarkable property, that distinguishes the QCD axion from {\emph{all}} other solutions to the strong CP problem. This feature is loved in particular by model-builders because thanks to it they are liberated from a number of otherwise annoying technical hurdles: the origin of the flavor hierarchy, the smallness of the weak scale, baryogenesis, dark matter, etc., are all questions that, in the presence of the QCD axion, may be addressed without the need to worry about possibly large CP-odd phases in the new physics. Nevertheless, (ii) is not at all an essential feature of a viable solution of the strong CP. One may in fact turn the argument upside down and argue that (ii) represents actually a ``drawback" of the QCD axion, since it indicates that even in the optimistic event of a confirmation of such a mechanism we will not be able to infer anything new about the nature of CP violation at short distances.

Leaving any philosophical drift aside, the reason $(iii)$ for its popularity is a very solid one. 
Any interesting scenario for physics beyond the SM must be able to make unambiguous predictions to be confronted with experimental data. The axion solution of the strong CP problem clearly predicts the existence of a pseudo-Nambu-Goldstone boson with an irreducible coupling to gluons
\ba
{\cal L}_{\rm axion}\supset\frac{\bar a}{f_a}\,\frac{g_{\rm C}^2}{32\pi^2}G^a_{\mu\nu}\widetilde G^{a\,\mu\nu}
\ea
where $\widetilde G^{a\,\mu\nu}=\epsilon^{\mu\nu\alpha\beta}G^a_{\alpha\beta}/2$, that can be tested via collider, astrophysical, as well as cosmological observations. Current data set limits of the order $f_a\gtrsim10^8-10^9$ GeV, leaving plenty of room for a solution of the strong CP problem. What is more is that a Bose condensate of very weakly-coupled axions may well be the totality or part of the dark matter if $f_a\sim10^{12}$ GeV.

Features (i), (ii), and (iii) are at the origin of the well-deserved popularity of the QCD axion. And yet, this nice story does not appear fully convincing to many of us. The reason is also well-known, so much so that it has been given a name: ``axion quality problem"~\cite{Kamionkowski:1992mf,Holman:1992us,Barr:1992qq}. The potential of the QCD axion is so small and the current lower bounds on $f_a$ so stringent that tiny perturbations from uncontrollable sources of $U(1)_{\rm PQ}$-breaking beyond the QCD anomaly, even if suppressed by the Planck scale, can easily destabilize the solution. In simple terms, the standard axion solution is apparently a very delicate one.

There are two paths to address this problem. One can either find a mechanism to suppress the potentially dangerous Planckian perturbations, or find new corrections to the potential such that the QCD contribution gets effectively enhanced. In either case, unfortunately, one or both of the attractive properties (i) and (ii) are lost. Known quantum field theories that address the quality problem by suggesting mechanisms to suppress quantum gravity perturbations invoke a number of seemingly ad-hoc gauge or global symmetries and some end up being rather baroque. This approach to the quality problem relies on such an intricate structure at short distances that the conceptual simplicity (i) promised by the QCD axion gets partially if not completely overshadowed. One may still hope that string theory offers alternative and more attractive explanations of the axion quality. Unfortunately, our current understanding of string theory makes it hard to firmly establish the relevance of this assertion. 

Rather than explaining why Planck-scale perturbations are suppressed compared to the QCD-induced potential, one may alternatively build a model with a heavy axion, in which the QCD potential is replaced by a much larger and more stable one. If we follow this path also the IR-effectiveness of the standard QCD axion, i.e. feature (ii) above, cannot hold anymore. In this paper we investigate scenarios that pursue this avenue. Here the challenge is to identify a framework in which the new $U(1)_{\rm PQ}$-breaking effects are perfectly aligned with the QCD-induced potential so that the overall energy is still minimized at a value $|\langle\bar a\rangle|/f_a\lesssim10^{-10}$, compatibly with current data. A few such mechanisms have been proposed.

One option is realized in scenarios where the large axion potential still comes from QCD but, as opposed to the standard mechanism, it is due to new short-distance effects~\cite{Holdom:1982ex}. If the QCD coupling grows relatively large at some UV scale, indeed, small instantons may become relevant, and such effects are naturally aligned with the low energy QCD potential. Unfortunately, a strongly-coupled UV framework of this type is intrinsically sensitive to the misaligning effect of whatever new CP-odd phases are present at the UV cutoff~\cite{Dine:1986bg,Kitano:2021fdl}. To firmly establish the viability of this program it is thus necessary to analyze a concrete realization. To the best of our knowledge the only explicit and tractable model of this type is the one of Ref. \cite{Agrawal:2017ksf}. This work shows that under reasonable assumptions the strong CP problem may in fact be solved, though the required setup introduces a few copies of the color gauge group along with a corresponding axion for each copy, and is therefore not as minimal as one might have hoped.

Another viable avenue is to postulate a scenario in which $U(1)_{\rm PQ}$ is anomalous under an additional non-abelian group ${\rm C}'$. Provided the anomaly coefficient is the same as the one of QCD, the axion coupling to the two sectors reads:
\ba\label{aggag'g'}
{\cal L}_{\rm axion}\supset\left(\bar\theta_{\rm C}+\frac{a}{f_a}\right)\frac{g_{\rm C}^2}{32\pi^2} G^a_{\mu\nu}\widetilde G^{a\,\mu\nu}+\left(\bar\theta_{\rm C'}+\frac{a}{f_a}\right)\frac{g_{\rm C'}^2}{32\pi^2}{G}^{\prime a}_{\mu\nu}{\widetilde G}^{\prime a\,\mu\nu},
\ea
with ${G'}^a_{\mu\nu}$ indicating the field strength of the ${\rm C}'$ vectors. If it is possible to further identify a structural condition that ensures 
\ba\label{theta33p}
\bar\theta_{\rm C}=\bar\theta_{\rm C'}
\ea
up to corrections smaller than $10^{-10}$, then a unique axion $\bar a/f_a=\bar\theta_{\rm C}+a/f_a$ can be defined. Its potential may be naturally dominated by the ${\rm C}'$ dynamics and be such that $\langle\bar a\rangle=0$, analogously to QCD~\cite{Vafa:1983tf,Vafa:1984xg}. In this framework one can obtain a sizable axion potential if the new non-abelian sector becomes strong at scales much larger than $\Lambda_{\rm QCD}$, and the quality problem is improved. The non-trivial task is explaining \eqref{theta33p}. 

We may justify \eqref{theta33p} by invoking a $Z_2$ symmetry~\cite{Rubakov:1997vp}. To realize this program a full copy of the SM is however needed, and in particular the new confining group must be a mirror copy of QCD, i.e. ${\rm C}'=SU(3)_{\rm C}'$. The mirror symmetry must be softly broken in order to ensure that the mirror sector be sufficiently heavy to have escaped detection. If the soft breaking is achieved via CP- and flavor-conserving interactions, any possible  correction to \eqref{theta33p} is controlled by loops of the SM Yukawas and higher-dimensional operators. It is known~\cite{Ellis:1978hq,Khriplovich:1985jr,Khriplovich:1993pf} that the former corrections, including both threshold as well as RG effects, are extremely small. The latter can be taken under control as well provided the soft breaking scale is sufficiently small compared to the UV cutoff. These mirror models are currently the most studied incarnation of heavy axion models~\cite{Berezhiani:2000gh,Hook:2014cda,Fukuda:2015ana,Dimopoulos:2016lvn,Hook:2019qoh}.

Yet, there may be a simpler and more minimal way to justify \eqref{theta33p}, one that does not require invoking a discrete mirror symmetry. One may in fact embed color $SU(3)_{\rm C}$ into a larger group at short distances, which we call {\emph{Grand Color}, and then postulate the latter be broken into the SM times the new confining group ${\rm C}'$. In this setup the structure of eq. \eqref{aggag'g'} emerges at the symmetry-breaking threshold, with \eqref{theta33p} easily satisfied at tree-level even when ${\rm C}'$ is not an $SU(3)$. This class of models was first suggested in Ref. \cite{Dimopoulos:1979pp}. However that paper does not present a concrete realization. Explicit models have been recently proposed in~\cite{Gherghetta:2016fhp,Gaillard:2018xgk}, but differ qualitatively from ours and the one of \cite{Dimopoulos:1979pp} due to the presence of mass terms for the exotic fermions. These latter make it difficult to ensure \eqref{theta33p} remains protected against radiative effects. The main results of the present paper are providing an explicit realization of the Grand Color scenario that robustly satisfies \eqref{theta33p} and presenting a careful study of the vacuum structure of the theory.

In Section \ref{sec:themodel} we introduce a model with ${\rm C}'=Sp(N-3)$ and argue that in such a framework the condition \eqref{theta33p} is satisfied up to negligible radiative effects and higher-dimensional operators, very much like in mirror models. The axion potential is analyzed in detail and proved to be aligned with the QCD one in Section \ref{sec:potential}. This is a highly non-trivial result because in our scenario the presence of Yukawa interactions prevents from applying the theorems of~\cite{Vafa:1983tf,Vafa:1984xg}. A large confinement scale for ${\rm C}'$ implies a sizable attenuation of the axion quality problem. The phenomenology is discussed in Section \ref{sec:phenomenology} and finally Section \ref{sec:conclusions} presents our conclusions.

\section{A Grand Color group}
\label{sec:themodel}

The gauge group of our model is $SU(N)_{\text{GC}}\times SU(2)_{\rm L}\times U(1)_{\rm Y'}$ and the entire matter content, SM included, is reported in table \ref{tab:GCmatter}. The Grand Color is an $SU(N)_{\text{GC}}$ gauge group and the SM quarks are in the fundamental and anti-fundamental representations. In order to cancel gauge anomalies, hypercharge must be partly embedded into the Grand Color and an abelian factor $U(1)_{Y'}$, while to avoid triviality $N$ must be odd~\cite{Witten:1982fp}. Yet, the leptonic sector remains basically the same as in the SM, whereas the scalar sector must include at least two additional fields, $\Phi$ in the adjoint and $\Xi$ in the 2-index anti-symmetric of $SU(N)_{\text{GC}}$, in order to break Grand Color in a phenomenologically viable way.

\begin{table}[h!]
	\begin{center}
		\renewcommand{\arraystretch}{1.2}
		\begin{tabular}{c|ccc}
			& $SU(N)_{\text{GC}}$            & $SU(2)_{\text{L}}$ & $U(1)_{\text{Y}'}$   \\
			\midrule
			$Q$                   & ${\bf N}$               & ${\bf 2}$ & $\frac{1}{2N}$\\
			$U$                & $\overline{\bf N}$ & ${\bf 1}$ & $-\frac{1}{2}-\frac{1}{2N}$ \\
			$D$                 & $\overline{\bf N}$ & ${\bf 1}$ & $+\frac{1}{2}-\frac{1}{2N}$\\
			$\ell$                  & ${\bf 1}$                & ${\bf 2}$ & $-\frac{1}{2}$ \\
			$e$                 & ${\bf 1}$               & ${\bf 1}$ & $+1$ \\
			\hline
			$H$                    & ${\bf 1}$                & ${\bf 2}$ & $+\frac{1}{2}$ \\
			$\Phi$  & ${\bf Adj}$             & ${\bf 1}$ & $0$ \\
			$\Xi$                  & ${\bf N}\otimes_A{\bf N}$     & ${\bf 1}$ & $\frac{1}{N}$ \\
			\bottomrule
		\end{tabular}
		\caption{Minimal field content of the model. The scalars $\Phi, \Xi$ are solely needed in order to break Grand Color into the SM gauge group.
		}\label{tab:GCmatter}
	\end{center}
\end{table}

The most general renormalizable Lagrangian for the fields in table \ref{tab:GCmatter} includes the standard kinetic terms and topological angles, a scalar potential, and a Yukawa interaction with the Higgs doublet $H$ of the same form as in the SM,
\begin{align}
	\mathcal{L}_{\text{Yuk}} = Y_u\, Q H U + Y_d \, Q \widetilde H D+ Y_e \, \ell  \widetilde H e+{\rm hc},
	\label{eq:Yukawas}
\end{align}
plus the operators $QQ \Xi \daga$ and $UD \Xi$. As explained in more detail below, though, the presence of the latter interactions would spoil the key relation \eqref{theta33p}. These couplings can be forbidden in several ways, for example gauging $\rm B-L$, promoting $\Xi$ to a composite scalar, or --- perhaps less elegantly --- invoking a global symmetry. Which of these mechanisms is actually at work does not concern us. In the following we will simply assume that \eqref{eq:Yukawas} represent the full set of renormalizable Yukawa interactions in our model.

The breaking of Grand Color is obtained in two steps:
\begin{eqnarray}
	SU(N)_{\text{GC}}\times SU(2)_{\text{L}} \times U(1)_{\text{Y}'}&\xrightarrow{\braket{\Phi}} & SU(3)_{\text{C}} \times SU(N-3)\times SU(2)_{\text{L}} \times U(1)_{\text{Y}'} \times U(1)_{\text{GC}} \nonumber \\
	&\xrightarrow{\braket{\Xi}} & SU(3)_{\text{C}} \times Sp(N-3)\times SU(2)_{\text{L}} \times U(1)_{\text{Y}}.
	\label{eq:breakingPattern}
\end{eqnarray}
In the first step the vev of a scalar $\Phi$ breaks $SU(N)_{\rm GC}$ into $SU(3)_{\text{C}} \times SU(N-3)\times U(1)_{\text{GC}}$. The abelian factor is normalized such that the fundamental representation of $SU(N)_{\text{GC}}$ decomposes as
\begin{align}\label{decomp}
\mathbf{N}\rightarrow (\mathbf{3},\mathbf{1})_{\frac{1}{6}-\frac{1}{2N}} \oplus (\mathbf{1},\mathbf{N-3})_{-\frac{1}{2N}}.
\end{align}
The second step consists in breaking $SU(N-3)\times U(1)_{\rm GC}\times U(1)_{\rm Y'} \xrightarrow{\braket{\Xi}} Sp(N-3)\times U(1)_{\rm Y}$ through the vev of the $\Xi$ component in the antisymmetric of $SU(N-3)$, which according to \eqref{decomp} carries a $U(1)_{\rm GC}$ charge equal to $-1/N$.~\footnote{To avoid any confusion, by $Sp(N-3)$ we denote the group of symplectic unitary $N-3$ matrices. Consistently, non-triviality of the theory implies $N-3$ is even.} 
It follows that the unbroken $U(1)_{\rm Y}$ charges are the sum of $U(1)_{\rm Y'}$ and the $U(1)_{\rm GC}$ generators. For simplicity we take both scalar vevs of order $f_{\rm GC}$. Importantly, because none of the new scalars $\Phi,\Xi$ has Yukawa couplings one can in principle promote both of them to composite operators. In that case there would no hidden fine-tuning in requiring the Grand Color breaking scale be much smaller than the UV cutoff, i.e. $f_{\rm GC}\ll f_{\rm UV}$. Strictly speaking, the only naturalness problem that our model necessary suffers from are the usual hierarchy and cosmological constant problems of the SM.

Below the Grand Color breaking scale $f_{\rm GC}$, the fields $Q,U,D$ split into the direct sum of the SM quarks plus exotic chiral fermions as shown in table \ref{tab:GCbreakingMatter}. The exotic fermions $\psi_{q,u,d}$ inherit the Yukawa couplings to $H$ from \eqref{eq:Yukawas} and are therefore formally the same as the SM ones up to renormalization effects. Crucially, however, there is no interaction between the SM fermions and the $\psi$'s apart from higher-dimensional operators suppressed by $f_{\rm GC}$. This implies that the flavor symmetries of the two sectors are effectively distinct: loops of the $Sp(N-3)$-charged sector will never be able to induce flavor-violating processes in the SM.

\begin{table}[h!]
	\begin{center}
		\renewcommand{\arraystretch}{1.3}
		\begin{tabular}{c|cccc}
			& $SU(3)_{\text{C}}$  &  $Sp(N-3)$  & $SU(2)_{\text{L}}$  & $U(1)_{\text{Y}}$   \\
			\midrule
			\multirow{2}{*}{$ Q = \begin{pmatrix} q \\ \psi_q \end{pmatrix} $} &$\mathbf{3}$&$\mathbf{1}$&$\mathbf{2}$& $\frac{1}{6} $\\
			& $\mathbf{1}$ & $\mathbf{N-3}$ & $\mathbf{2}$ & $0$\\
			\multirow{2}{*}{$ U = \begin{pmatrix} u \\ \psi_u \end{pmatrix} $} &$\mathbf{\bar 3}$&$\mathbf{1}$&$\mathbf{1}$& $-\frac{2}{3}$ \\
			& $\mathbf{1}$ & $\mathbf{N-3}$ & $\mathbf{1}$ & $-\frac{1}{2}$ \\
			\multirow{2}{*}{$ D = \begin{pmatrix} d \\ \psi_d \end{pmatrix} $} &$\mathbf{\bar 3}$&$\mathbf{1}$&$\mathbf{1}$& $\frac{1}{3}$ \\
			& $\mathbf{1}$ & $\mathbf{N-3}$ & $\mathbf{1}$ & $\frac{1}{2}$ \\
			\bottomrule
		\end{tabular}
		\caption{Decomposition of the quarks below the scale $f_{\rm GC}$. Here $\psi_q=(\psi_{q_u},\psi_{q_d})$ is an electroweak doublet. The SM hypercharge $U(1)_{\text{Y}}$ is the sum of $U(1)_{\text{Y}'}$ and $U(1)_ {\text{GC}}\subset SU(N)_ {\text{GC}}$.}\label{tab:GCbreakingMatter}
	\end{center}
\end{table}

In addition, the field-basis invariant $SU(3)_{\rm C}$ and $Sp(N-3)$ topological angles, inherited by Grand Color as shown in \eqref{aggag'g'}, at tree-level satisfy $\bar\theta_{\rm C}=\bar\theta_{\rm C'}= \theta-\arg\det Y_u Y_d$, where $\theta$ denotes the $SU(N)_{\rm GC}$ angle. Radiative effects can spoil this tree-level relation, and it is mandatory for us to show that the misaligning affects are under control. There are three different sources of radiative effects that can potentially invalidate  \eqref{theta33p}: the scalar sector, the Yukawa couplings, and non-renormalizable interactions. The vevs of $\Phi$ and $\Xi$ are the order parameters of Grand Color breaking and their insertion is necessary to generate a difference in the two topological angles. Other than that, however, the scalar sector cannot appreciably contribute to a violation of \eqref{theta33p} since the most general renormalizable potential $V(H,\Phi,\Xi)$ is automatically CP-conserving and its parameters can always be chosen so that CP does not get broken spontaneously. Furthermore, all radiative corrections due to the Yukawa sector \eqref{eq:Yukawas} at and below $f_{\rm GC}$ are known to be completely negligible~\cite{Ellis:1978hq,Khriplovich:1985jr,Khriplovich:1993pf}. Had we allowed the presence of unsuppressed flavor-violating coefficients for $QQ \Xi \daga$, $UD \Xi$, this nice property would not have held anymore.\footnote{
In this respect our approach differs qualitatively from~\cite{Gherghetta:2016fhp,Gaillard:2018xgk}, where the beyond the SM fermions filling the Grand Color multiplet are decoupled by giving them large masses. Such a decoupling may also be achieved in our model, where $QQ \Xi \daga$, $UD \Xi$ would generate a vector-like mass matrix $M$ for the $Sp(N-3)$ fermions below $f_{\rm GC}$. Unfortunately, decoupling would typically violate \eqref{theta33p} because $M$ introduces a new physical CP-odd phase that contributes to ${\bar\theta}_{\rm C'}$ at tree-level. In order to preserve $|{\bar\theta}_{\rm C'}-{\bar\theta}_{\rm C}|<10^{-10}$ one would therefore be forced to demand that $|{\rm Arg}[{\rm det}[M]]|<10^{-10}$. In this paper we avoid this fine-tuning by forbidding the couplings $QQ \Xi \daga$, $UD \Xi$. This way the extra fermions remain chiral, like the SM fermions, and get trapped into the heavy $Sp(N-3)$ hadrons.
}
}

The bottom line is that in our model eq.\eqref{theta33p} remains satisfied up to the desired accuracy at the renormalizable level. The most dangerous non-renormalizable interactions are dimension-5 and dimension-6 operators that contribute differently to the topological angles once the scalars $\Phi,\Xi$ acquire a vev:
\begin{align}\label{UVcutoffPhi}
	\frac{\bar c_5}{f_{\text{UV}}} \frac{g_{\rm GC}^2}{32\pi^2}\Phi \, G_{\rm GC}\tilde G_{\rm GC}, \; \frac{\bar c_6}{f_{\text{UV}}^2} \frac{g_{\rm GC}^2}{32\pi^2}\Phi\daga \Phi \,  G_{\rm GC}\tilde G_{\rm GC}, \; \frac{\bar c_6'}{f_{\text{UV}}^2} \frac{g_{\rm GC}^2}{32\pi^2}\Xi\daga \Xi \,  G_{\rm GC}\tilde G_{\rm GC}.
\end{align}
The dominant effect comes from the first interaction, but this can be avoided by charging $\Phi$ under an additional gauge symmetry, or postulating that $\Phi$ be the scalar responsible for breaking $U(1)_{\rm PQ}$, in which case $f_a\sim f_{\rm GC}$. The last two operators are more model-independent and imply $\bar\theta_{\rm C}-\bar\theta_{\rm C'}\sim f_{\rm GC}^2/f_{\rm UV}^2$. Taking the Planck scale $f_{\text{UV}}=2.4\times10^{18}$ GeV as the UV cutoff, satisfying the relation eq.\eqref{theta33p} up to corrections of order $10^{-10}$ imposes the constraint $f_{\text{GC}} \lesssim 10^{13}\gev$. This bound can be further relaxed if $\Phi,\Xi$ are composite operators.

Overall, the picture that emerges is qualitatively similar to the $Z_2$-symmetric scenarios: color and the exotic confining dynamics have basically the same topological angle if no Yukawa couplings are introduced beyond $Y_u,Y_d$ and the breaking of Grand Color is sufficiently soft. Under these conditions a unique axion $\bar a/f_a=\bar\theta_{\rm C}+a/f_a$ from the breaking of a $U(1)_{\rm PQ}$ with a Grand Color anomaly would automatically relax to zero the topological angles of both color and ${\rm C'}=Sp(N-3)$. By making the latter confine at a scale $f\gg f_\pi$ much larger than QCD we will see the axion mass can be enhanced and the axion quality improved while still robustly solving the strong CP problem. Actually, we will have to require $f$  larger than the weak scale because the exotic fermions carry electroweak charges (see also Section \ref{sec:phenomenology}).

The precise origin of the axion is not relevant to our work. What matters is that its couplings to the $SU(3)_{\rm C}\times Sp(N-3)$ topological terms be the same. In addition, we will work under the hypothesis that 
\ba\label{regimefaf}
f_a>f, 
\ea
so that the tools of effective field theory can be employed in the next section to study the axion potential.~\footnote{The opposite regime, with $f_a<f$, may nevertheless provide a solution to the strong CP problem but requires a completely different study.} Scenarios in which $f_a>f_{\rm GC}$ automatically lead to equal couplings to the $SU(3)_{\rm C}\times Sp(N-3)$ topological terms. It is perhaps worth showing explicitly that the same may also be true for $f_a<f_{\rm GC}$. To see this let us for example UV complete the axion sector via an interaction $\mathcal{L}_{\text{PQ}} \supset y F F^c \Theta$, with $F$ ($F^c$) fermions in the fundamental (anti-fundamental) of $SU(N)_{\text{GC}}$ carrying $U(1)_{\rm PQ}$ charge $+1$ and $\Theta$ a scalar of charge $-2$ responsible for breaking $U(1)_{\rm PQ}$ spontaneously at a scale $\sim f_a$. In such a model the axion acquires the very same couplings to the $SU(3)_{\rm C}\times Sp(N-3)$ topological terms even with $f_a<f_{\rm GC}$ because below the Grand Color breaking scale $F,F^c$ split into the direct sum of fermions that are both in the fundamental representation of $SU(3)_{\rm C}$ and $Sp(N-3)$ and so have the same Dynkin index. The phase in $y$ does not affect this conclusion. Finally, we note that for definiteness we decided to work within a KSVZ axion model, but it should be clear that a DFSZ model would equally do.

\section{The axion potential}
\label{sec:potential}

At scales below $f_{\rm GC}$ our model reduces to the SM plus an $Sp(N-3)$ gauge theory with three families of fermions $\psi_q=(\psi_{q_u},\psi_{q_d})$, $\psi_{u,d}$ charged as shown in table \ref{tab:GCbreakingMatter}, with Yukawa couplings \eqref{eq:Yukawas}, and an axion $\bar a$ equally coupled to color $SU(3)_{\rm C}$ and $Sp(N-3)$. All the scalars contained in $\Phi,\Xi$ acquire masses proportional to $f_{\rm GC}$ and decouple. In this section we discuss the fate of the exotic fermions and the axion potential.

The basic assumption is that $Sp(N-3)$ confines at a scale $f<f_{\rm GC}$ larger than $v\approx246$ GeV. This hypothesis is certainly realized provided $N\geq9$. In order to get rid of an otherwise large mixing between the axion and the $\eta'$ singlet of the $Sp(N-3)$ dynamics we remove the axion from the topological term via a rotation of the $\psi_{q,u,d}$. This can for example be achieved via a phase re-definition $\psi_u\to e^{i\bar a/{3f_a}}\psi_u$, where the factor of $3$ denotes the number of generations, which puts the axion in front of the up-type Yukawas, i.e. $Y_{u}\to e^{i\bar a/{3f_a}}Y_{u}$. 

The physics at confinement is better described in terms of a strong $Sp(N-3)$ dynamics with an approximate $SU(12)$ global symmetry under which the column vector $\Psi = (\psi_{q_u}, \psi_{q_d}, \psi_u, \psi_d)$ transforms as the fundamental representation. At confinement the chiral condensates $\langle\psi_{q_u}\psi_{q_d}\rangle=-\langle\psi_{q_d}\psi_{q_u}\rangle=\langle\psi_{u}\psi_{d}\rangle\sim4\pi f^3/\sqrt{N}$ break $SU(12)\to Sp(12)$~\cite{Peskin:1980gc,Preskill:1980mz}. To demonstrate this we first observe that, because all bound states of $Sp(N-3)$ are bosonic, 't Hooft anomaly matching implies that $SU(12)$ must be broken. Finally, by the Vafa-Witten theorem we know that the vectorial subgroup, namely $Sp(12)$, should remain unbroken~\cite{Kosower:1984aw}. Crucially, the electroweak symmetry is part of the unbroken group. The choice ${\rm C'}=Sp(N-3)$ is essential to achieve this key property.

The pattern $SU(12)\to Sp(12)$ delivers 65 would-be Nambu-Goldstone bosons (NGBs) $\Pi$. These are not exact because the weak gauging of $SU(2)_{\rm L}\times U(1)_{\rm Y}$ and the Yukawa couplings constitute a small explicit breaking of $SU(12)$. In particular 51 of the would-be NGBs acquire positive mass squared of order $g^2f^2,g'^2f^2$ from loops of the $SU(2)_{\rm L}\times U(1)_{\rm Y}$ vectors. The other 14, denoted by $\Pi_0$, are gauge-neutral and can in principle mix with $\bar a$, similarly to the $\pi_0$ in the standard QCD axion. It is the dynamics of these $\Pi_0$ that controls vacuum alignment and in particular the vacuum expectation value of the axion. The vev of the charged NGBs, instead, vanish and can be ignored in our discussion. Incidentally, some of the charged NGBs are electroweak doublets and mix with the fundamental $H$. The heavy linear combinations are integrated out, whereas we assume that the mass parameter of $H$ is such that there exists a unique light eigenstate with a small and negative mass squared. This will play the role of the Higgs doublet of the SM, $H_{\rm SM}$. The fine-tuning we just invoked is nothing but the usual hierarchy problem.~\footnote{Note that from this observation follows that the true SM Yukawa couplings in low-energy observables differ compared to $Y_{u,d}$, not only due to different RG effects, but also because of some mixing angle. We will neglect these corrections since our results are anyway affected by uncertainties of ${\cal O}(1)$ from incalculable coefficients.}

The dynamics of the neutral NGBs can be effectively described observing that the electroweak symmetry leaves intact a smaller $SU(3)_q \times SU(3)_u \times SU(3)_d \times U(1)_{\text{B}}$ global subgroup of $SU(12)$, associated to the independent flavor rotations of $\psi_q,\psi_u,\psi_d$ as well as the $Sp(N-3)$ baryon number under which $\psi_u,\psi_d$ have charge opposite to $\psi_q$. The vacuum condensates break this symmetry down to $SO(3)_q \times SU(3)_{u-d}$. As a result the 14 neutral NGBs can be effectively parametrized in terms of three matrices: a special, unitary and symmetric matrix $\Sigma_L\in SU(3)_q /SO(3)_q$, a special, unitary matrix $\Sigma_R\in SU(3)_u \times SU(3)_d/ SU(3)_{u-d}$, and finally $\eta_{\rm B}$, the NGB of $U(1)_{\rm B}$. The boson $\eta_{\rm B}$ remains an exactly massless state because the baryon number is not explicitly broken. We will discuss its phenomenology in Section \ref{sec:phenomenology}. The remaining 13 neutral scalars, along with the axion, acquire a potential from the Yukawa interactions of $\psi_{q,u,d}$. We stress that these couplings are the same as those of the SM quarks at the threshold $f_{\rm GC}$, though below that scale they renormalize differently. At the scale $f$ relevant for the present discussion the $\psi_{q,u,d}$ couplings, which to avoid over-complicating our notation will still be denoted by $Y_{u,d}$, are expected to be somewhat larger than the SM Yukawas by a flavor-universal factor due to loops of the $Sp(N-3)$ dynamics. Expanding in powers of $Y_{u,d}$ the most general potential reads:
\begin{align}
\label{eq:potential}
	V_{\text{neutral}} = \frac{c_{ud}}{N}  f^4 \, \tr \left[Y_u \Sigma_R Y_d ^t \Sigma_L \right] e^{i \frac{\bar a}{N_gf_a}} + {\rm hc}+{\cal O}(Y^4,v^2/f^2),
\end{align}
with $N_g=3$ the number of generations. The dominant contribution arises from a loop of $H$ and the $Sp(N-3)$ dynamics. The factor of $N$ has been identified using a large $N$ scaling and recalling that $f^2\propto N$. The parameter $c_{ud}$ is a real incalculable quantity. Subleading corrections contain $|H_{\rm SM}|^2$ and/or higher order insertions of the Yukawa couplings. The former cannot affect qualitatively the potential; such corrections are necessarily small because we are interested in the chiral regime $v\lesssim f$ (see also Section \ref{sec:phenomenology}). The latter will be argued to be negligible in Section \ref{sec:strongCP}.

Contrary to the standard QCD axion, the theory under consideration contains a light fundamental scalar with Yukawa couplings to the fermions $\Psi$ and it is not possible to directly apply the results of~\cite{Vafa:1984xg} in order to argue that $\langle\bar a\rangle=0$. The minimization problem is therefore conceptually different from QCD. In particular, in QCD~\cite{Vafa:1984xg} imply that the vev of the pions must vanish and the low energy dynamics contains no CP violation other than the one encoded in $\bar\theta$. In our scenario, on the other hand, the axion effective potential can depend non-trivially on the vacuum configuration of the $\Pi_0$'s. We will have to prove $\langle\bar a\rangle=0$ by brute force. This is what we will do in the next section.

Before turning to the minimization of the potential, though, we stress that \eqref{eq:potential} possesses a $Z_{N_g} \subset SU(4 N_g)$ symmetry under which $\Sigma_{R,L} \rightarrow e^{\pm i 2 \pi n /N_g } \Sigma_{R,L}$. This discrete symmetry signals the presence of a set of inequivalent vacua sharing the same perturbative mass spectrum and axion vev, which may indicate a cosmological domain-wall problem if the temperature of the Universe ever exceeded $f$.~\footnote{These domain-walls are stable because $Z_3\subset U(1)_{\rm B}$, and explicit breaking of $U(1)_{\rm B}$ occurs via effective operators of an extremely high dimensionality, since the baryon number is very well protected by our gauge symmetries.} This issue adds to the more familiar domain-wall problem of axion models, which takes place at the scale $f_a>f$.

\subsection{Minimisation at leading order}
\label{subsec:minimisation}

The potential (\ref{eq:potential}) involves 14 fields (13 neutral NGBs $\Pi_0$ and the axion $\bar a$), and its minimization is highly non-trivial. To perform this task we find it convenient to first discuss the properties of the more general structure
\begin{align}\label{pot.gen}
	V_{\text{neutral}}^{\rm LO} = V_0 \left(\Pi_0/f\right)e^{i\bar a/f_a N_g} + {\rm hc},
\end{align}
where the number of light fermion generations $N_g$ as well as the explicit expression of $V_0$ are left arbitrary. Interestingly, both the potentials of our model and that of the standard QCD axion have precisely this form. Therefore some of the results discussed here have a rather general validity. In particular, in appendix \ref{app:potential} we demonstrate that the absolute minimum of \eqref{pot.gen} is found by maximizing $|V_0|$, whereas the axion vacuum is determined by $\braket{\bar a}/f_a=N_g (\pi-\phi ) \text{ mod } 2\pi$, where $\phi={\rm arg}[V_0]$ at the extremum.

In the case at hand $V_0$ is given in equation (\ref{eq:potential}), and in the basis in which $Y_u = \widehat Y_u$ is diagonal can be written as
\begin{eqnarray}
\label{eq:V0general}
	V_0
	&=& \frac{c_{ud}}{N}  f^4 \, \tr [\widehat Y_u\Sigma_R\widehat Y_d V_{\text{CKM}}\daga\Sigma_L]\\\no
	&=& \frac{c_{ud}}{N}  f^4 \,  [\widehat Y_u]_i[A]_{ii},~~~~~~~~~~~A=\Sigma_R\widehat Y_d V_{\rm CKM} \daga \Sigma_L.
\end{eqnarray}
$|V_0|$ is maximized when $A$ is aligned as much as possible along $\widehat Y_u$, with the corresponding entries satisfying $|[A]_{33}|>|[A]_{22}|>|[A]_{11}|$. Suppose for the time being that it is possible to find a configuration $\Sigma_{L,R}$ that fully diagonalizes $A$, so that the diagonal entries read $[A]_{ii}=|[A]_{ii}|e^{i\phi_i}$, where $\phi_i$ are phases subject to $\phi_1+\phi_2+\phi_3=2\pi n$ because of the constraint $\det [A]=\det [\widehat Y_d] \in \mathds{R}$. Under this hypothesis $|V_0|$ would be maximized when $\phi_i=\phi_j$ is common to all entries, such that the trace becomes a coherent sum of terms, and the minimum configuration would read $\phi_i=2\pi n/N_g$. The phase of $V_0$ at the minimum would finally be $\phi=\arg \, c_{ud}+2\pi n/N_g$, and from eq. (\ref{eq:App_VacV0}) we would infer that $\braket{\bar a}/f_a=N_g(\pi-(\arg \, c_{ud}+2\pi n/N_g))=N_g(\pi- \arg \, c_{ud})$, or
\begin{eqnarray}
\label{eq:axionvev}
\frac{\braket{\bar a}}{f_a}=
\begin{cases} 
	0 \text{ mod }2\pi & {\rm if}~N_g={\rm even}\\
	0 \text{ mod }2\pi & {\rm if}~N_g={\rm odd~and}~c_{ud}<0\\
	\pi \text{ mod }2\pi & {\rm if}~N_g={\rm odd~and}~c_{ud}>0
\end{cases}.
\end{eqnarray}
This shows that, as long as $A$ can be diagonalized, the system has a natural tendency to relax the axion to a CP-conserving vev. Thus the axion vev is usually vanishing, though for odd $N_g$ and positive $c_{ud}$ we get ${\braket{\bar a}}/{f_a}=\pi$. Despite being CP-conserving, the latter option is not phenomenologically acceptable because incompatible with the Gell-Mann-Okubo relations~\cite{Crewther:1979pi}. In the standard QCD axion the result of~\cite{Vafa:1984xg} ensures that ${\braket{\bar a}}/{f_a}=0$, which implies that $c_{ud}$ must be negative. In our model later on we will offer some argument indicating that $c_{ud}$ should be negative.

Unfortunately, it is possible to prove that as soon as $N_g\geq3$ the matrix $A$ cannot be exactly diagonalized because $\Sigma_L$, being unitary-symmetric, does not contain enough degrees of freedom to diagonalize $A^\dagger A$. In scenarios with $N_g\geq3$ the logic leading to \eqref{eq:axionvev} can thus at most be approximate. And yet, we find (at least for the physically relevant case $N_g=3$) that \eqref{eq:axionvev} remains valid. Despite the impossibility of diagonalizing $A$, in fact, the relation ${\rm det}[A]=A_{11}A_{22}A_{33}+\Delta$ holds up to a very small perturbation $\Delta$. Eq. \eqref{eq:axionvev} then applies because in a perturbative expansion for small off-diagonal elements the dynamical phases of the three diagonal elements of $A$ are determined at leading order to be $2\pi n/N_g$. That is, the corresponding fluctuations fall into a deep potential well, which cannot be destabilized by the next to leading corrections due to $\Delta$. As a result the overall phase of $V_0$ is still determined by $\phi=\arg \, c_{ud}+2\pi n/N_g$ and the axion vev by \eqref{eq:axionvev}, as if $A$ could be exactly diagonalized.

Even though the above arguments seem rather convincing to us, an explicit calculation would help lifting any doubt on \eqref{eq:axionvev}. Furthermore, an explicit analysis is necessary to compute the masses of the NGBs and the axion. In the following we will thus verify \eqref{eq:axionvev} and calculate the axion mass for $N_g=1$, where in fact $A$ is trivially diagonalized, as well as for $N_g=2$, where it can be fully diagonalized by the NGB matrices. Subsequently we will consider the phenomenologically relevant case $N_g=3$. Along the way we will argue in favor of $c_{ud}<0$.

\paragraph{Warming up with $N_g=1$ and $N_g=2$} \mbox{}\\

The $N_g=1$ case is almost trivial, since the spectrum of NGBs is composed of a charged composite Higgs, that is not relevant to vacuum alignment, and the exact flat direction $\eta_{\rm B}$. The potential simply reduces to a potential for the axion:
\begin{align}\label{VNg=1}
	V_{\text{neutral}}^{\rm LO}  =  2 \, \frac{c_{ud}}{N}  f^4 \,  y_u y_d \cos\left(\frac{\bar a}{f_a}\right)  \qquad \qquad \qquad (N_g=1)
\end{align}
where $y_u,y_d$ are the up and down quark Yukawas. This potential is minimised at $\braket{\bar a}/f_a=0\text{ mod }2\pi$ if $c_{ud}<0$ or $\braket{\bar a}/f_a=\pi\text{ mod }2\pi$ if $c_{ud}>0$, as expected from \eqref{eq:axionvev}. The axion mass is given by
\begin{align}\label{massNg=1}
	m_a ^2 = 2 \, \frac{|c_{ud}|}{N} \,  y_u y_d \frac{f^4}{f_a^2}   \qquad (N_g=1).
\end{align}
It is possible to show that for $N_g=1$ the parameter $c_{ud}$ must be negative. The argument is a bit involved and will only be sketched here. 

Our argument starts by considering a modified $N_g=1$ scenario in which $Y_u=Y_d$ and only the neutral component of the fundamental Higgs is dynamical. This is certainly not our model, but its effective potential is just a simple generalization of ours because the UV diagrams that generate it, in terms of fundamental fermions and $H$, are virtually identical to those in our model modulo corrections of order $v^2/f^2$. In particular, the sign of the overall coefficient $c_{ud}$ is exactly the same in the two scenarios because determined by equal correlators in the unperturbed $Sp(N-3)$ theory. The conclusion that the axion vev vanishes only for $c_{ud}<0$ remains valid. But crucially, in the modified model the fermionic determinant arising from the integration of $\Psi$ is real and positive definite because the fermionic spectrum is effectively doubled, see \cite{Vafa:1983tf}. Therefore the result of~\cite{Vafa:1984xg} can be used to argue that the axion must be minimized at zero, and hence indirectly that $c_{ud}<0$. This for us is proof that the coefficient $c_{ud}$ in \eqref{VNg=1} is negative.

The minimization of the $N_g=2$ case is more interesting. In this case the potential \eqref{eq:potential} depends on a CKM matrix that can be written in terms of the Cabibbo angle:
\begin{align}
	V_{\text{CKM}} = \begin{pmatrix}
		\cos \theta_c & \sin \theta_c\\
		-\sin \theta_c & \cos \theta_c
	\end{pmatrix}  \qquad (N_g=2).
\end{align}
As anticipated at the end of Section \ref{sec:potential}, we find two inequivalent vacua related by a $Z_2$ symmetry. These are given by
\begin{eqnarray}\label{Ng=2sol}
	c_{ud}>0:~~~
	\langle\Sigma_{L}\rangle=
	(\pm)\left(\begin{matrix}
		i\cos\theta_c & i\sin\theta_c\\
		i\sin\theta_c & -i\cos\theta_c
	\end{matrix}\right),
	~~~~~~~~
	\langle\Sigma_{R}\rangle=
	(\pm)\left(\begin{matrix}
		i & 0\\
		0 & -i
	\end{matrix}\right),~~~~~~\braket{\bar a}=0, \\\label{Ng=2sol'}
	c_{ud}<0:~~~
	\langle\Sigma_{L}\rangle=
	(\pm)\left(\begin{matrix}
		i\cos\theta_c & i\sin\theta_c\\
		i\sin\theta_c & -i\cos\theta_c
	\end{matrix}\right),
	~~~~~~~~
	\langle\Sigma_{R}\rangle=
	(\mp)\left(\begin{matrix}
		i & 0\\
		0 & -i
	\end{matrix}\right),~~~~~~\braket{\bar a}=0.
\end{eqnarray}
With two generations the vacuum configurations precisely diagonalise $A =\Sigma_R \widehat Y_d V_{\text{CKM}}\daga \Sigma_L$, and in both cases $\braket{\bar a}/f_a=0$ consistently with \eqref{eq:axionvev}. In this vacuum all scalar excitations (except for the exact flat direction $\eta_{\rm B}$) are massive. 

The $N_g=2$ case is so simple to handle analytically that we were able to find an explicit expression for the effective axion potential. This is obtained by solving the equation of motion for the neutral NGBs $\Pi_0$ and plugging it back into $V_{\rm neutral}$. It is a reliable approximation of the axion self-couplings in the limit $f\ll f_a$ in which the neutral NGBs are much heavier than the axion. We find
\begin{align}\label{potNg=2}
	V_{\text{eff}} \, \left(\frac{\bar a}{f_a} \right)= - 2\, \frac{|c_{ud}|}{N}  f^4 \, \tr [\widehat Y_u\widehat Y_d]\sqrt{1-4\frac{\det[\widehat Y_u \widehat Y_d]}{\left(\tr [\widehat Y_u\widehat Y_d] \right)^2}\sin^2 \left(\frac{\bar a}{2f_a}\right) } \qquad (N_g=2),
\end{align}
which is consistently minimized at $\langle\bar a\rangle/f_a=0$ mod $2\pi$. This result is reminiscent of the potential of the QCD axion in 2-flavor QCD. The axion mass immediately follows:
\begin{align}\label{massNg=2}
	m_a ^2 = 2 \frac{|c_{ud}|}{N}   \, \frac{\det[\widehat Y_u \widehat Y_d]}{\tr [\widehat Y_u\widehat Y_d] } \, \frac{f^4}{f_a^2}  \qquad (N_g=2).
\end{align}
Eq. \eqref{potNg=2} is very valuable to us because we will not be able to obtain an explicit expression for $N_g=3$. It is therefore useful to extract as much information as possible from it. First, we observe that in the limit of a heavy second generation eqs \eqref{potNg=2}, \eqref{massNg=2} reduce to eqs \eqref{VNg=1}, \eqref{massNg=1}. This is a highly non-trivial check of the consistency of our results. It is a consequence of the fact that a large Yukawa coupling for the second generation implies that a number of NGBs becomes much heavier than those associated to the light first generation. Up to corrections suppressed by the heavy NGB mass, therefore, the potential should reduce to the $N_g=1$ case, which is what we see here explicitly. The very same logic constrains the structure of the $N_g=3$ potential, as we will verify numerically. 

A second important lesson we can draw from \eqref{potNg=2}, and more readily \eqref{massNg=1}, is that the non-trivial dependence of the axion potential should be controlled by $\det[\widehat Y_u \widehat Y_d]$. This is also a very general result, independent of $N_g$. Indeed, if any of the eigenvalues of $Y_u$ or $Y_d$ were to vanish the UV Lagrangian would be invariant under an additional anomalous axial symmetry which could be combined with $U(1)_{\rm PQ}$ to obtain an exact unbroken one. In that situation the axion would become an exact flat direction. Hence the axion potential must be proportional to at least a power of all the eigenvalues. The quantity $\det[\widehat Y_u \widehat Y_d]$ is the simplest object with this property. The axion mass squared cannot be simply proportional to the determinant unless $N_g=1$, however. By counting the units of $\hbar$ we need at least $2N_g-2$ additional coupling constants in the denominator, so dimensional analysis forces $m_a^2$ to be inversely proportional to an appropriate combination of $Y_u,Y_d$ as found in \eqref{massNg=2}. The significant hierarchy in the SM fermion masses and the decoupling properties mentioned in the previous paragraph, together indicate that the axion mass in our model is always numerically close to \eqref{massNg=1}.

\paragraph{The real world: $N_g=3$} \mbox{}\\

Having checked the simplified scenarios $N_g=1,2$, we can now turn to the realistic case $N_g=3$. The previous calculations support the correctness of our qualitative argument of section \ref{subsec:minimisation}, and as such we expect the leading order potential \eqref{eq:potential} to be minimised at $\braket{\bar a}/f_a = 0\,  (\pi)$ for $c_{ud}<0 \, (>0)$. Unfortunately, in the case $N_g=3$ the potential involves 14 fields and it is not possible to approach the problem analytically. For this reason we employ a customised {\small \sc MATHEMATICA} algorithm which enables us to numerically find the minimum of the potential up to a very high accuracy. The minimisation procedure is repeated many times in order to statistically validate the result. The Yukawa couplings that appear in the potential are renormalized at the scale $\sim4\pi f/\sqrt{N}$ by the $Sp(N-3)$ dynamics. As a benchmark we employ the PDG data for $V_{\text{CKM}}$ and the numerical values of $\widehat Y_u, \widehat Y_d$ that correspond to the SM quark Yukawas evaluated at the TeV scale. Changing the numerical value of these couplings does not affect our results qualitatively. What we find is exactly what anticipated in section \ref{subsec:minimisation}: the distinct vacua configurations are related by a $Z_3$ symmetry $\Sigma_{R,L}\rightarrow e^{\pm i 2\pi n/3}\Sigma_{R,L}$; all vacua give rise to the same $A$, which is diagonal up to small off-diagonal elements; the axion is minimised at $\braket{\bar a}/f_a = 0$ or $\pi$ depending on the sign of $c_{ud}$.

In the study of the $N_g=1$ toy model we gave an argument supporting the claim that $c_{ud}$ is negative. At sufficiently large $N$ there is no distinction between the coefficients $c_{ud}$ for the $N_g=1$ and $N_g=3$ scenarios. We are therefore motivated to conjecture that $c_{ud}<0$ also for $N_g=3$, and from now on work under this hypothesis. Given the central role played by this hypothesis, it would be interesting to find an independent proof, for instance using lattice QCD techniques.

We are now interested in studying the spectrum of the NGBs and the axion. To read off the masses we first canonically normalise the kinetic term $f^2{\rm tr}[\partial_\mu\Sigma_{L,R}\partial^\mu\Sigma_{L,R}]$, altered by the vev of the NGBs, and then perform an $SO(14)$ rotation to diagonalise the Hessian of the potential. As a result of this operation, the masses of the 13 neutral NGBs are found to be
\begin{align}\label{NGBmasses}
	m_{\Pi_0}^2 \simeq  \left(
	\begin{matrix} 
	6.4\times 10^{-2}\\
	2.5\times 10^{-2}\\
	2.4\times 10^{-2}\\
	2.4\times 10^{-2}\\
	2.4\times 10^{-2}\\
	2.4\times 10^{-2}\\
	2.3\times 10^{-2}\\
	2.7 \times 10^{-5}\\
	1.2\times 10^{-5}\\
	2.5\times 10^{-6}\\
	1.6\times 10^{-6}\\
	1.6\times 10^{-6}\\
	4.0\times 10^{-8}	
	\end{matrix}\right) \times \frac{|c_{ud}|}{N}  f^2
\end{align}
for all the three $Z_3$-symmetric vacua configurations. The eigenstate corresponding to the axion is the lightest one for any $f<f_a$. The mass that we extract numerically respects the scaling suggested in the previous section, namely
\begin{align}\label{eq:axionmassma}
	 m_a ^2 \simeq 2 \frac{|c_{ud}|}{N}   y_uy_d \, \frac{f^4}{f_a^2}  \simeq1.8\times  10^{-10} \times \frac{|c_{ud}|}{N} \, \frac{f^4}{f_a^2} \qquad (N_g=3).
\end{align}
The NGBs' masses are substantially unaffected by the mixing with the axion as long as $f/f_a \ll 1$ (even though the mixing will turn out to be important for phenomenology, see section \ref{sec:phenomenology}). In the extreme limit $f/f_a\to1$ the mixing impacts the NGBs masses by a few $10\%$, with the lightest being affected the most.

We conclude this section emphasizing that the axion mass \eqref{eq:axionmassma} is parametrically enhanced with respect to the standard QCD axion as long as $f\gtrsim 10^2  f_{\pi}  $. The goal outlined in Section \ref{sec:introduction} is fulfilled.

\subsection{Subleading corrections and heavy axion quality}
\label{sec:strongCP}

The analysis carried out so far demonstrates that our axion has a large mass and a leading order potential minimized at $\braket{\bar a}=0$. Higher order corrections cannot affect this result unless they introduce new sizable sources of CP-violation or flavor-violation. We will argue next that subleading effects due to renormalizable interactions do not spoil our solution of the strong CP problem and that the effect of non-renormalizable operators can be taken under control. 

\paragraph{Renormalizable interactions} \mbox{}\\

In the renormalizable version of our scenario the effective axion potential $V_{\rm eff}(\bar a/f_a)$ depends on flavor-invariant combinations of the parameters $Y_{u,d}$, $\langle\Sigma_{L,R}\rangle$. The Yukawas parametrize {\emph{explicit}} CP violation, whereas the NGB vacuum potentially represents an independent source of {\emph{spontaneous}} CP violation. A non-vanishing vev for the axion is induced by CP-odd, flavor-conserving combinations of these parameters. 

Crucially, in our model all CP-odd invariants must necessarily be proportional to explicit CP-violation. This follows from the fact that spontaneous CP violation does not take place. As a first simple check of this statement, let us inspect the $N_g=2$ toy model, where we have an explicit analytic solution. Here the CKM matrix is real, i.e. there is no explicit CP violation, and it is readily seen that eqs \eqref{Ng=2sol} and \eqref{Ng=2sol'} preserve the generalised CP transformation $\Sigma_{R,L} \to -\Sigma_{R,L}^*$. A far less trivial check is obtained for $N_g=3$. In that case we verified that, when we switch off the CKM phase, the explicit numerical solution of the leading order potential in \eqref{eq:potential} also satisfies the relation  $\braket{\Sigma_{R,L}} = e^{\pm i 2\pi n/3}\braket{\Sigma_{R,L}}^*$. In other words, in the absence of explicit violation, CP is not spontaneously broken.

An important consequence of what we just demonstrated is that any CP-odd flavor invariant in $V_{\rm eff}$ must be proportional to the explicit CP-violation in $Y_{u,d}$. A key property of our theory, inherited from the SM, is that explicit CP-violation should disappear whenever two of the eigenvalues of the SM quark mass matrix squared are degenerate, or any mixing angle goes to zero, or when the CKM phase vanishes. This is very important. We have already seen that the non-trivial part of the effective axion potential must be proportional to ${\rm det}[Y_uY_d]$. Here we find that any explicit CP-violating interaction of the axion must contain a further suppression that disappears in the above limits. Such a suppression is so significant that explicit CP-violation in $V_{\rm eff}$ becomes effectively innocuos: CP-odd flavor-invariant combinations of the Yukawas and $\langle\Sigma_{L,R}\rangle$ may arise in $V_{\rm eff}$ only at very high order in an expansion in $Y_{u,d}$ and are numerically extremely small. It is therefore not surprising that, even including the CKM phase, our ${\cal O}(Y^2)$ potential does not induce an axion vev. In fact, at ${\cal O}(Y^2)$ the effect of explicit CP violation cannot be visible, the only flavor-invariant candidate $\tr \left[Y_u \braket{\Sigma_R} Y_d ^t \braket{\Sigma_L} \right]$ is real and hence there is nothing that can be on the right-hand side of $\braket{\bar a}=0$.

In summary, subleading corrections to the effective axion potential are either even in $\bar a$ or odd, the latter being proportional to the explicit CP-violation. Terms even in $\bar a$ cannot destabilize our solution because the leading order theory has no flat directions. The terms odd in $\bar a$ are however dangerous if they include a tadpole. In that case the axion vev is shifted from the origin. Still, the shift must be proportional to the tiny explicit CP-violating phase that controls the tadpole and the vacuum expectation value would thus be safely below $|\langle{\bar a}\rangle|/f_a\lesssim10^{-10}$. Our axion dynamically solves the strong CP problem like in the standard QCD scenario. Higher dimensional operators with new CP-violating or flavor-violating couplings can however introduce new CP-odd flavor invariants which can be numerically more relevant that those of the renormalizable theory. These effects are discussed next.

\paragraph{Higher-dimensional operators} \mbox{}\\

At the root of the axion quality problem is the fact that a huge $f_a$ makes the axion potential extremely sensitive to cutoff-suppressed $U(1)_{\rm PQ}$-breaking interactions. As reviewed in Section \ref{sec:introduction} this sensitivity may be alleviated by increasing the axion mass. However, there is no free lunch. To enhance the axion mass, the confinement scale $f$ has to be rather large as well. As a result, cutoff-suppressed $U(1)_{\rm PQ}$-{\emph{conserving}} contributions to the axion potential, as long as they are CP- or flavor-violating, may become important and in principle spoil the solution of the strong CP problem.

More precisely, consider the flavor-violating but $U(1)_{\rm PQ}$-preserving operator 
\ba\label{UVVaxion}
	\frac{\bar c_{ijkl}}{f_{\text{UV}} ^2} Q_iQ_jU_kD_l,
\ea
where $i,j,k,l$ are flavor indices. This has precisely the same axial $U(1)_A$ charges as the leading order potential in \eqref{eq:potential} and thus represents a modification $\delta V_0\sim\bar c\,16\pi^2f^6/(N^2f_{\rm UV}^2)$ of the quantity $V_0\sim {\rm Tr}[Y_uY_d] f^4/N$ defined in \eqref{eq:V0general}. This is not aligned with \eqref{eq:potential} for generic ${\bar c}_{ijkl}$, and must therefore be small. According to eq.\eqref{axionsolgeneralappendix} the axion vev is of order $\langle\bar a\rangle/f_a\sim{\rm Im}[\delta V_0]/|V_0|$, where the imaginary part of the flavor invariant $\delta V_0$ may come either from ${\bar c}_{ijkl}$ directly or from phases of the leading NGB vev, which become physical when contracted with a flavor-violating ${\bar c}_{ijkl}$. The requirement that the effective topological angle be less than $10^{-10}$ becomes
\ba\label{boundfUV}
{f}\lesssim10^{-7}f_{\rm UV},
\ea
which for a maximal UV cutoff of order $f_{\rm UV}=2.4\times10^{18}~{\rm GeV}$ reads ${f}\lesssim10^{11}$ GeV. Somewhat similar considerations apply to operators that do not violate the axial $Sp(N-3)$ symmetry $U(1)_A$ but still violate flavor, like
\ba
\frac{\bar c_{ijkl}}{f_{\rm UV}^2}(\Psi_i\Psi_j)(\Psi_k\Psi_l)^\dagger.
\ea
This operator modifies the NGB vev and in turn shifts the axion minimum. As a conservative bound we impose \eqref{boundfUV}. Moreover, we could have operators that do not violate flavor, but contribute new CP-odd flavor-conserving phases to the axion effective potential. A typical example is the Weinberg operator
\ba
\frac{\bar c_W}{M_{\rm UV}^2}\frac{g^3_{\rm GC}}{16\pi^2}G_{\rm GC}G_{\rm GC}\widetilde G_{\rm GC}.
\ea
The new CP-odd phase should be smaller than $10^{-10}$ to guarantee a solution of the strong CP problem. From this requirement the weaker upper bound $\bar c_Wf^2/f^2_{\rm UV}\lesssim10^{-10}$ follows.

Finally, let us come back to the original motivation of our work: the axion quality problem. As a crude estimate, imposing that a $U(1)_{\rm PQ}$-violating operator of dimension $d$ does not significantly alter the axion vev implies
\begin{eqnarray}
	g_{\rm UV}^2 f_a ^4\left(\frac{f_a}{f_{\text{UV}}}\right)^{d-4}\lesssim 10^{-10}m_a^2f_a^2,
\end{eqnarray}
where $g_{\rm UV}$ is a coupling of the UV dynamics. For a given value of $f_a$ this represents a lower bound on $d-4$. Comparing to the lower bound in the standard QCD axion, i.e. $(d-4)_{\text{std}}$, this reads
\ba\label{eq:PQquality}
(d-4)=(d-4)_{\text{std}}-\frac{\ln{m_a^2}/{m_{a,{\rm std}}^2}}{\ln{f_{\rm UV}}/{f_a}}.
\ea
The quality problem is logarithmically sensitive to the mass ratio ${m_a^2}/{m_{a,{\rm std}}^2}$ and becomes more and more sensitive to this quantity as $f_a$ gets larger. In our model we find an appreciable improvement as long as $f\gg10^2f_\pi$, see \eqref{eq:axionmassma}. As a numerical example, for $f_a=10^{10}$ GeV and $f = 10^8$ GeV we get an axion mass of a few GeV. This corresponds to $d \gtrsim 7$, which is a significant improvement compared to $d_{\text{std}}\gtrsim 10$.

\section{Phenomenology}
\label{sec:phenomenology}

The phenomenology of our model is extremely rich and cannot be investigated in depth here. In this section we present a first qualitative assessment. 

The $Sp(N-3)$ dynamics generates many massive hadrons, all of which are unstable because there is no unbroken flavor symmetry that protects them. Heavy hadrons of mass $\propto4\pi f/\sqrt{N}$ as well as baryons quickly decay into NGBs. The electroweak-charged NGBs decay into the SM Higgs boson, $W^\pm,Z^0$, and neutral $\Pi_0$'s. The latter are much more long-lived, and decay dominantly into QCD hadrons and/or photons via the mixing with the axion and the $\eta'$ of the $Sp(N-3)$ dynamics. Less relevant decay channels for $\Pi_0$'s are into SM fermions via non-renormalizable interactions generated at the scale $f_{\rm GC}$. Very likely, yet, only the lightest hadrons were significantly produced in the early Universe because in order to robustly avoid a domain-wall problem associated to the $Z_3$-degeneracy of the NGB potential reheating must probably have occurred after $Sp(N-3)$ confinement (see discussion below eq. \eqref{eq:potential}).

The hadrons can be directly produced at the LHC and future colliders. In addition, the $Sp(N-3)$ dynamics can be indirectly probed via precision measurements. As a rough measure of the current impact of these constraints we impose the qualitative bound $f\gtrsim$ TeV. In this regime the low energy signatures are mainly controlled by $\bar a$ and $\Pi_0$ (and, if present, $\eta_B$; see below). The effective field theory is governed by the couplings to the topological terms of the gluon and the photon
\ba\label{EFT}
	\lagr_{\rm EFT} &\supset& \frac{1}{2}(\partial\bar a)^2-\frac{m_a^2}{2}\bar a^2+\frac{g_{\rm C} ^2}{32\pi^2} \frac{\bar a}{f_a} G\widetilde G + \bar c_{a\gamma\gamma} \frac{e ^2}{32\pi^2} \frac{\bar a}{f_a} F\widetilde F\\\no
	&+&\frac{1}{2}(\partial\Pi_{0,i})^2-\frac{m_{\Pi_0,i}^2}{2}\Pi_{0,i}^2+\frac{g_{\rm C} ^2}{32\pi^2} \bar c_{\Pi_0gg}^i\frac{\Pi_{0,i}}{f} G\widetilde G + \bar c^i_{\Pi_0\gamma\gamma} \frac{e^2}{32\pi^2} \frac{\Pi_{0,i}}{f} F\widetilde F.
\ea
The effective couplings to the $Z^0$ and $W^\pm$ bosons are phenomenologically less relevant. Assuming that the $U(1)_{\rm PQ}$ has no electroweak anomaly, and momentarily ignoring the mixing with $\Pi_0$, the coefficient $\bar c_{a\gamma\gamma}$ can be computed by moving the axion from the $SU(N)_{\text{GC}}$ topological term to the Yukawas with an anomalous chiral rotation, and re-placing it only in front of the QCD topological term below the Grand Color breaking. Recalling that the SM hypercharge is given by a combination of $U(1)_{\text{Y}'}$ and $U(1)_{\text{GC}}$, we get
\begin{align}\label{cagammagamma}
	\bar c_{a\gamma\gamma} 
	=-\frac{1}{2}\left(N-3\right)\left(1-\frac{1}{3N}\right).
\end{align}
Consistently with expectations, this expression vanishes for $N=3$, when our model reduces to a standard KSVZ scenario. We will consider $N=13$ for definiteness, noting that a number of colors $>17$ would typically induce a Landau pole for $SU(2)_{\rm L}$ below the Planck scale whereas for $N<9$ the condition $f\gtrsim$ TeV would not be attained. We verified that with these parameters the condition $f<f_{\rm GC} \lesssim 10^{13}\gev$ is also satisfied.

The neutral NGBs have no bare coupling to the SM vectors. However, they acquire them from the mixing with $\bar a$ and the heavy $\eta'$ of $Sp(N-3)$. These contributions are parametrically of order $\bar c_{\Pi_0gg}\sim f/f_a$ and $\bar c_{\Pi_0\gamma\gamma}\sim{\rm max}[m_{\Pi_0}^2/m_{\eta'}^2,f/f_a]$. As a result the decay rate into gluons, controlled by the $\bar a-\Pi_0$ mixing, is of order
\ba\label{decayAxion}
\Gamma_{\Pi_0\to gg}\gtrsim\Gamma_{\bar a\to gg}\frac{f_a}{f},
\ea
where we took into account the different scaling of the masses with $f,f_a$ (see \eqref{eq:axionmassma} and \eqref{NGBmasses}). Importantly, the rate is always greater than $\Gamma_{\bar a\to gg}$ in the regime \eqref{regimefaf}. This parametric estimate is confirmed by an accurate numerical analysis, which also reveals that some $\Pi_0$ can have rates several orders of magnitude larger than shown in \eqref{decayAxion}. Note also that the NGB mixing with the axion does not appreciably modify \eqref{cagammagamma}. As long as $f\ll f_a$ the effect is parametrically suppressed, and we numerically verified that as $f$ approaches $f_a$ the change in the axion coupling to photons is still at most ${\cal O}(10\%)$.

\begin{figure}[!t]
\begin{center}
	\includegraphics[width = 1\textwidth]{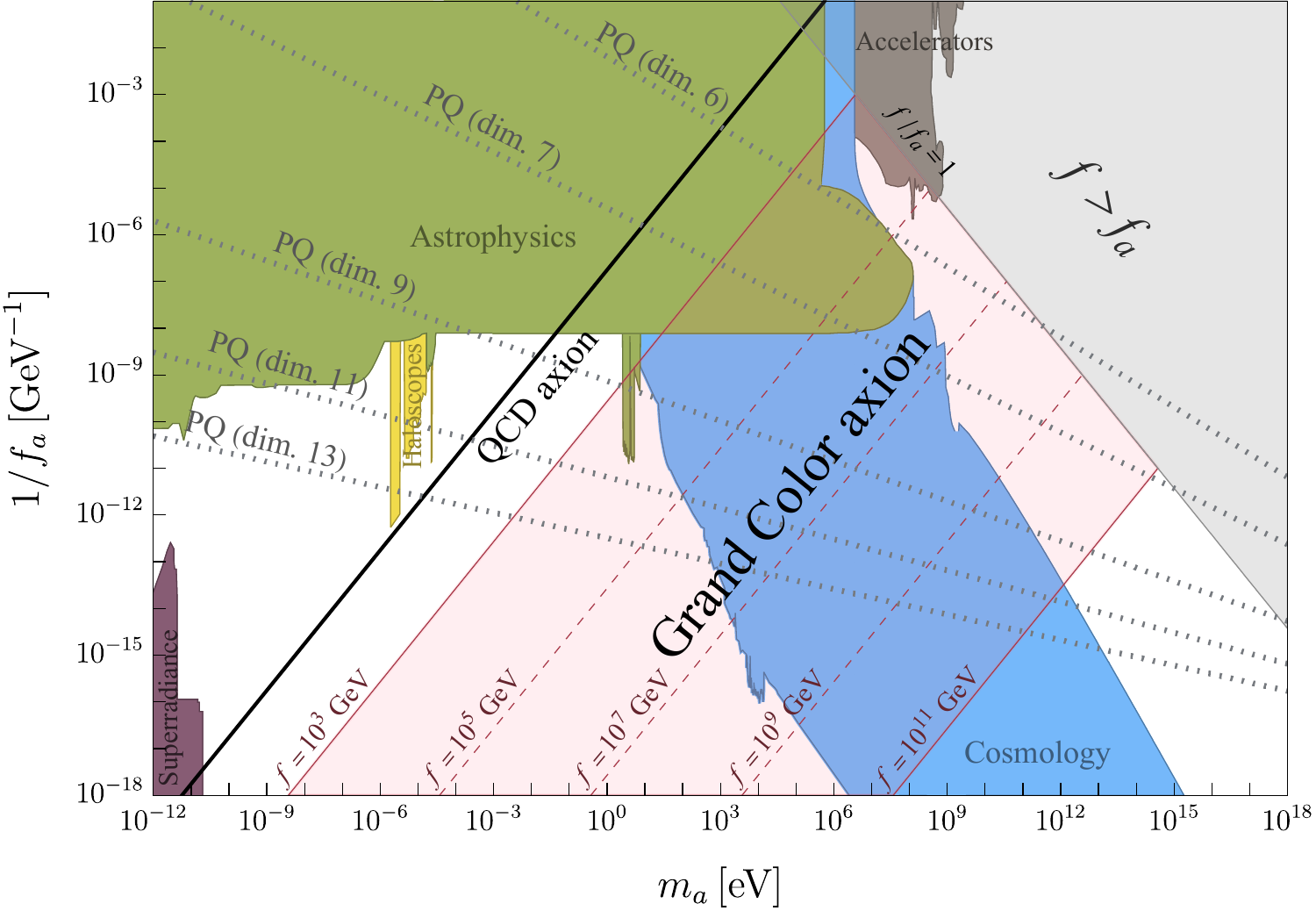}
	\caption{Collection of the main bounds on $f_a$ vs $m_a$, as discussed in section \ref{sec:phenomenology}. For definiteness we assumed $N=13$ and $|c_{ud}|=1$. The light-pink bend denoted by ``Grand Color axion" identifies the region populated by our scenario.}
	\label{fig:axionBounds1}
	\end{center}
\end{figure}

The axion mass is given in eq. \eqref{eq:axionmassma} and in the allowed regime $f\gtrsim$ TeV is always larger than the standard one. Furthermore, as we saw around \eqref{boundfUV}, a conservative condition for the strong CP problem to be solved is $f\lesssim10^{11}$ GeV, where we identified the UV cutoff with the Planck scale. Combining the two bounds we see that our scenario populates the light-pink bend in the $m_a-f_a$ plot of Fig. \ref{fig:axionBounds1} labeled by ``Grand Color axion", defined by the implicit relation $10^3~{\rm GeV}\leq f\leq10^{11}$ GeV --- where $f=f(m_a,f_a)$ is given by \eqref{eq:axionmassma}. We included a hard cut at $f<f_a$ to indicate the regime of validity of the effective field theory approach adopted in this paper, see eq. \eqref{regimefaf}. In the grey region $f>f_a$ our results do not necessarily apply, though without a detailed analysis this region cannot be excluded. The dotted grey lines in Fig. \ref{fig:axionBounds1} show contour regions of eq. \eqref{eq:PQquality} with $f_{\rm UV}=2.4\times10^{18}$ GeV. The axion quality problem is progressively more alleviated as we move towards the upper-right corner. The regime with high quality is the one in which, for a fixed $f_a$, the axion mass is maximal, i.e. the confinement scale reaches the extreme value $f\to f_a$ compatible with \eqref{regimefaf}.

The ``Grand Color axion" bend is mostly probed by cosmological observations from BBN and CMB physics (blue), astrophysics (green), and collider experiments (brown). In particular, the left boundary of the blue region is taken from the collection of bounds in \cite{AxionLimits}. The rightmost part of the cosmology bound is due mainly to the $N_{\rm eff}$ bounds from \cite{Dunsky:2022uoq} and the requirement that the total axion decay rate satisfies $\Gamma_{\rm tot}\geq3H(T_{\rm BBN})$, where $H(T_{\rm BBN})$ is the Hubble rate when the Universe reached temperatures of order $T_{\rm BBN}=4$ MeV \cite{deSalas:2015glj}, in order not to interfere with Big-Bang-Nucleosynthesis. The hadronic decay rate has been calculated adapting the results of \cite{Aloni:2018vki} to our model. The axion decay rate takes into account \eqref{cagammagamma}, includes also the mixing with the QCD mesons, and is modified with respect to the standard case because of the much larger value of $m_a$. For the astrophysics bounds we refer to \cite{AxionLimits}. The collider bounds on top of Fig. \ref{fig:axionBounds1} are taken from \cite{Goudzovski:2022vbt} and \cite{Kelly:2020dda}. The white regions are currently viable. 

In the far bottom-left of the ``Grand Color axion" bend one should make sure that the heavier NGBs are sufficiently long-lived to avoid disrupting the primordial abundance of light elements. However, because of their large masses, vacuum misalignment typically over-produces them in the form of a Bose condensate of $\bar a,\Pi_0$ unless the initial misalignment angles are extremely small. Moreover, in that region the axion quality problem is not ameliorated compared to the standard scenario. For these reasons we believe the upper-right region is more motivated.

The upper-right corner of the ``Grand Color axion" bend is also more interesting because in that regime ongoing and future accelerators as well as future CMB observations are able to explore our model. In particular, because $f_a$ can approach the TeV scale this scenario can be probed for example at  HL-LHC \cite{Hook:2019qoh}, Kaon and Hyperon factories \cite{Goudzovski:2022vbt}, DUNE \cite{Kelly:2020dda}, NA62 \cite{Ertas:2020xcc}, Belle II \cite{Chakraborty:2021wda,Bertholet:2021hjl} and MATHUSLA \cite{Chou:2016lxi}. The phenomenological signatures are much richer than the standard QCD axion because of a variety of axion-like-particles $\Pi_0$ with anomalous couplings, see \eqref{EFT}. Future CMB surveys will also significantly improve measurements of $N_{\rm eff}$ and will be able to constrain a wider region of parameter space on the right of the blue area. In contrast to the accelerator searches mentioned above, however, here the neutral NGBs are not expected to play any relevant role because are much heavier and decay faster than the axion, see \eqref{decayAxion}. Yet, they might still lead to non-standard cosmological signatures, albeit quite indirect. The energy stored in a Bose condensate of neutral NGBs via vacuum misalignment may be estimated as $\rho_{\Pi_0}(T)\sim\theta_r^2m_{\Pi_0}^2f^2({T}/{T_r})^3$, with $\theta_r$ denoting the misalignment angle at the end of inflation and $T_r$ the reheating temperature. For certain values of masses and decay rates the temperature $T_{\rm m}$ at which $\rho_{\Pi_0}(T_{\rm m})$ dominates over radiation is actually larger than the decay temperature $T_{\rm d}\sim\sqrt{\Gamma_{\Pi_0} M_{\rm Pl}}$. When this happens the Universe undergoes an early period of matter domination which might result in a depletion of the primordial densities of visible and dark matter.

One last comment should be added before closing. In addition to $\Pi_0,\bar a$, our model generically features a virtually massless photophobic axion-like particle $\eta_B$ from the breaking of $U(1)_{\rm B}\subset SU(12)$. It acquires no potential from the Yukawa interactions, since these are $U(1)_{\rm B}$-symmetric, and only has electroweak anomalous couplings. Its effective Lagrangian reduces to
\ba\label{EFT1}
	\lagr_{\rm EFT} &\supset& \frac{1}{2}(\partial\eta_{\rm B})^2+\frac{\eta_{\rm B}}{f_B} \left(\frac{g_{\rm L}^2}{32\pi^2} W\widetilde W - \frac{g_{\rm Y}^2}{32\pi^2} Y\widetilde Y\right),
\ea
where $f_B=(2/\sqrt{3})f/N$, plus derivative interactions with itself and the other NGBs. While its coupling to photons vanish, 1-loop generated interactions to the SM fermions lead to a constraint of order $f\gtrsim 300\times N$ TeV~\cite{Craig:2018kne} whereas, under the hypothesis $T_f<f$, the impact on $N_{\rm eff}$ is minimal. The bound on $f$ just quoted is much stronger than those considered before. To obliterate the problem we may get rid of $\eta_B$ by gauging $U(1)_{\rm B-L}$, as suggested for other reasons in Section \ref{sec:themodel}. In order for $\eta_{\rm{B}}$ to be gauged away, all the fundamental scalars should be neutral and right-handed neutrinos should be added in order to ensure gauge anomalies cancellation.

\section{Conclusions}
\label{sec:conclusions}

We constructed a concrete model for a heavy axion arising from an enlarged color sector, which is conceptually and structurally very simple, and has an extremely minimal field content. Including all SM fields as well, the latter entirely fits in Table \ref{tab:GCmatter}. No mirror copy of the SM, nor of color, is required. In addition, there is no need for additional global symmetries besides the familiar $U(1)_{\rm PQ}$. The necessary coupling structure may naturally result as a consequence of gauge invariance and the field content. 

The unique sources of CP-violation in the renormalizable part of the theory are included in the Yukawa couplings $Y_{u,d}$ and the topological angles, exactly as in the SM. Hence, radiative corrections cannot spoil the equality of the couplings of the axion to QCD and the new confining group ${\rm C'}=Sp(N-3)$. From the very same reason follows that the effective Grand Color axion potential is automatically aligned with the QCD one: our scenario introduces no exotic flavor-conserving CP-odd phases, and is therefore approximately CP-invariant up to a very high accuracy, like the QCD potential. 

Our scalar potential has a characteristic ${\cal O}(Y^2)$ form but also shares some similarity with the standard one due to the presence of a mixing between the axion and the heavy neutral Nambu-Goldstone modes, somewhat analogous to the $\bar a-\pi_0$ mixing in the standard scenario. The resulting Grand Color axion mass is very distinctive and scales as:
\ba\label{AxionMassScaling}
m_a^2\sim \frac{y_uy_d}{N} \frac{f^4}{f_a^2},
\ea
where $y_{u,d}$ are the up- and down-quark Yukawas renormalized by the $Sp(N-3)$ dynamics at $\sim4\pi f/\sqrt{N}$. By construction the axion potential in our model is a 2-loop-sized effect proportional to two powers of Yukawas, and eq. \eqref{AxionMassScaling} follows from dimensional analysis and simple physical considerations. This expression differs from the one predicted by existing heavy axion models and its magnitude falls somewhat in between mirror and UV-instanton models. It is significantly enhanced compared to the one predicted by potentials dominated by small instantons. As a result, an improvement in axion quality is achieved with a significantly smaller $f$, and hence a reduced sensitivity to physics at the cutoff scale. The scaling in \eqref{AxionMassScaling} is however suppressed compared to what is found in $Z_2$-symmetric models, and so for a similar axion mass our $f$ needs to be larger.

While corrections to the effective topological angle $\langle\bar a\rangle/f_a$ from renormalizable couplings are virtually negligible, the effect of $U(1)_{\rm PQ}$-violating as well as $U(1)_{\rm PQ}$-conserving higher-dimensional operators can in principle alter the axion vev in a significant way. The sensitivity to Peccei-Quinn-violating interactions suppressed by the UV cutoff is reduced compared to the standard scenario, and the quality is certainly improved because of the larger axion mass. Yet, we found that heavy axion models develop a novel sensitivity to {\emph{Peccei-Quinn-preserving}} deformations.~\footnote{During the completion of this work, ref. \cite{Bedi:2022qrd} appeared stressing this very same point. Their numerical estimates are based on 1-instanton calculations and therefore not reliable for our model nor for mirror-symmetric scenarios. Yet their conclusions are qualitatively general and agree with ours.} This ``heavy axion quality problem" is generically shared by all models that attempt to increase the axion mass via a new strong coupling at $f\gg f_\pi$, and ours is no exception. Future measurements of the dipole moments of the neutron, as well as of atoms and molecules will potentially be able to set upper bounds on the axion mass of these scenarios.

Finally, we should point out that in this paper we presented just one of a larger set of interesting Grand Color scenarios. Our choice ${\rm C}'=Sp(N-3)$ was for example motivated by the need to avoid electroweak symmetry breaking at the scale $f$, but other options are possible. It would be interesting to explore alternative Grand Color scenarios, for example models with different ${\rm C}'$, scenarios compatible with a complete grand unification into a simple gauge group, or scenarios with composite $\Phi,\Xi$, and construct explicit models with $f_a\sim f$. Besides robustly addressing the strong CP problem with an alleviated quality problem, these models feature a number of distinctive phenomenological signatures, which in this paper we just began to explore. 

\section*{Acknowledgments}
\label{sec:acknowledgments}

We thank F. D'Eramo and L. Di Luzio for discussions on the cosmological and collider bounds. This research was partly supported by the Italian MIUR under contract 2017FMJFMW (PRIN2017), the “iniziativa specifica” Physics at the Energy, Intensity, and Astroparticle Frontiers (APINE) of Istituto Nazionale di Fisica Nucleare (INFN), and the European Union’s Horizon 2020 research and innovation programme under the Marie Skłodowska-Curie grant agreement No~860881-HIDDeN.

\appendix

\section{General considerations on the leading order potential}
\label{app:potential}

In this appendix we derive the minimization condition for the leading order potential of the neutral scalars. This can be compactly written as in eq. \eqref{pot.gen}. The extremality conditions read
\begin{align}
	\begin{cases}
		\frac{\delta V_0}{\delta\Pi_m}e^{i \braket{\bar a}/f_aN_g}+\frac{\delta V^*_0}{\delta\Pi_m}e^{-i\braket{\bar a}/f_aN_g}=0\\
		V_0e^{i\braket{\bar a}/f_aN_g}-V_0^*e^{-i\braket{\bar a}/f_aN_g}=0.
	\end{cases}
\end{align}
Combining these equations (we will show below that $|V_0| \neq 0$) we get $e^{i\braket{\bar a}/f_a N_g} = \pm V_0 ^* /|V_0|$, so that the above system can be rewritten as 
\begin{align}\label{axionsolgeneralappendix}
	\begin{cases}
		\frac{\delta |V_0|}{\delta\Pi_m}=0\\
		\sin \left( \frac{\braket{\bar a}}{N_g f_a} \right)=\mp\frac{{\rm Im}[V_0]}{|V_0|}
	\end{cases}
\end{align}
from which it is clear that the vacuum is obtained extremizing $|V_0|$, and a necessary condition for the strong CP problem to be solved is that $V_0$ is real at the minimum. 

To see whether the vacuum is actually at the minimum or at the maximum of $|V_0|$ we must study the Hessian, which reads
\begin{align}
	{\cal H}= \left(\begin{matrix}
		M_{\Pi\Pi}^2 & M^2_{a\Pi} \\
		M^2_{a\Pi} &  M_{aa}^2
	\end{matrix}
	\right)
\end{align}
with
\begin{eqnarray}
	[M_{\Pi\Pi} ^2]_{mn}&=& \frac{1}{f^2} \frac{\delta^2V_0}{\delta\Pi_m\delta\Pi_n}e^{i\braket{\bar a}/f_a N_g}+\frac{1}{f^2}\frac{\delta^2V^*_0}{\delta\Pi_m\delta\Pi_n}e^{-i\braket{\bar a}/f_a N_g}\\
	\left[M_{a\Pi} ^2 \right]_m  &=& \frac{i}{f f_aN_g}\frac{\delta V_0}{\delta\Pi_m}e^{i\braket{\bar a}/f_a N_g}-\frac{i}{f f_a N_g}\frac{\delta V^*_0}{\delta\Pi_m}e^{-i\braket{\bar a}/f_a N_g} \\
	M^2 _{aa}&=&-\frac{1}{f_a ^2 N^2_g}V_0e^{i\braket{\bar a}/f_a N_g}-\frac{1}{f_a^2 N^2_g}V^*_0e^{-i\braket{\bar a}/f_a N_g}.\nonumber
\end{eqnarray}
The NGB-axion mixing and the pure axion term are order $f/f_a$ and $f^2/f_a^2$. For simplicity we will work in the limit $f\ll f_a$, so we can treat them as perturbations, but our conclusions will apply in general. In this way, if $M^2_{\Pi\Pi}$ has no flat directions the lightest eigenvalue approximately reads
\begin{align}
	M_a^2=M_{aa}^2-M^2_{a\Pi}[M^2_{\Pi\Pi}]^{-1}M^2_{a\Pi}+{\cal O}(f^4/f_a^4)\leq M^2_{aa}.
	\label{eq:App_axionmass}
\end{align}
Thus a necessary condition for stability is $M_{aa}^2\geq M_{a}^2\geq0$, which translates into
\begin{align}
	\label{eq:App_VacV0}
	V_0e^{i\braket{\bar a}/N_g f_a}=-|V_0|.
\end{align}
This tells that the absolute minimum of the potential is reached when $V_{\text{neutral}}^{\rm LO}=2 V_0e^{i\braket{\bar a}/N_g f_a}=-2|V_0|$, and hence the vacuum is obtained by maximizing $|V_0|$. This justifies the earlier assumption $V_0\neq0$, for $V_0=0$ would be energetically disfavored. The conclusion just derived has been obtained for $f\ll f_a$ but in fact has general validity because a mass mixing always pushes the lightest eigenstate to lower values, so the condition $M_{aa}^2\geq0$ is anyway necessary. The presence of flat directions in $M^2_{\Pi\Pi}$ would not alter the conclusion either. Indeed, for the same reason we just explained these directions cannot mix with the axion otherwise they would turn tachyonic after the mixing is removed. It follows that $\Pi$ flat directions cannot affect the axion mass nor the argument leading to \eqref{eq:App_VacV0}.

In summary, we demonstrated in complete generality that the minimum of $V^{\rm LO}_{\text{neutral}}$ is obtained by maximizing $|V_0|$ with respect to the goldstone fields. The axion vev follows. Indeed, writing $V_0 = |V_0 (\braket{\Pi})|e^{i \phi ( \braket{\Pi}/f ) }$, eq. (\ref{eq:App_VacV0}) indicates that the value of the axion at the minimum is determined by $\braket{\bar a}/N_g f_a + \phi =\pi \text{ mod } 2\pi$.

The solution of the strong CP problem requires $N_g(\pi - \phi )$ be a multiple of $2\pi$, and this cannot be assessed unless an explicit form of $V_0$ is given. In Section \ref{subsec:minimisation} we analyzed in detail the potential of our scenario. Here, as a quick check of our results, we study the leading order potential for the standard axion in 2-flavor QCD. Note that the same structure \eqref{eq:V0general} applies to the standard QCD axion provided $\widehat Y_u$ is interpreted as the quark mass matrix and $A$ the pion matrix. In the QCD case the axion is rotated such that the quark masses are positive, i.e. $m_{u,d}>0$, and the axion field appears as in \eqref{pot.gen} with $N_g=2$. Switching off the charged pion components we have $V_0=C[m_ue^{i\pi_0/f_\pi}+m_de^{-i\pi_0/f_\pi}]$, with $C$ some constant. For $m_u\neq0$ we find that $|V_0|$ has two extrema, one at $\langle\pi_0\rangle=0$ and the other at $\langle\pi_0\rangle=\pi/2$. At the two extrema the function $|V_0|$ is respectively given by $m_u+m_d$ and $|m_u-m_d|$. Hence the absolute maximum of $|V_0|$ is attained when $\langle\pi_0\rangle=0$ and the strong CP problem, as well-known, is solved, i.e. $\braket{\bar a}/f_a = 2\pi\sim0$. When $m_u=0$ the function $|V_0|=|C|m_d$ is constant and $\phi={\rm arg}[V_0]$ arbitrary. The argument above tells us that the axion is now a flat direction, as expected.

\bibliography{biblio.bib}

\begin{thebibliography}{10}

\bibitem{Peccei:1977hh}
R.~D. Peccei and Helen~R. Quinn.
\newblock {CP Conservation in the Presence of Instantons}.
\newblock {\em Phys. Rev. Lett.}, 38:1440--1443, 1977.
\newblock \href {https://doi.org/10.1103/PhysRevLett.38.1440}
  {\path{doi:10.1103/PhysRevLett.38.1440}}.

\bibitem{Weinberg:1977ma}
Steven Weinberg.
\newblock {A New Light Boson?}
\newblock {\em Phys. Rev. Lett.}, 40:223--226, 1978.
\newblock \href {https://doi.org/10.1103/PhysRevLett.40.223}
  {\path{doi:10.1103/PhysRevLett.40.223}}.

\bibitem{Wilczek:1977pj}
Frank Wilczek.
\newblock {Problem of Strong $P$ and $T$ Invariance in the Presence of
  Instantons}.
\newblock {\em Phys. Rev. Lett.}, 40:279--282, 1978.
\newblock \href {https://doi.org/10.1103/PhysRevLett.40.279}
  {\path{doi:10.1103/PhysRevLett.40.279}}.

\bibitem{Kamionkowski:1992mf}
Marc Kamionkowski and John March-Russell.
\newblock {Planck scale physics and the Peccei-Quinn mechanism}.
\newblock {\em Phys. Lett. B}, 282:137--141, 1992.
\newblock \href {http://arxiv.org/abs/hep-th/9202003}
  {\path{arXiv:hep-th/9202003}}, \href
  {https://doi.org/10.1016/0370-2693(92)90492-M}
  {\path{doi:10.1016/0370-2693(92)90492-M}}.

\bibitem{Holman:1992us}
Richard Holman, Stephen D.~H. Hsu, Thomas~W. Kephart, Edward~W. Kolb, Richard
  Watkins, and Lawrence~M. Widrow.
\newblock {Solutions to the strong CP problem in a world with gravity}.
\newblock {\em Phys. Lett. B}, 282:132--136, 1992.
\newblock \href {http://arxiv.org/abs/hep-ph/9203206}
  {\path{arXiv:hep-ph/9203206}}, \href
  {https://doi.org/10.1016/0370-2693(92)90491-L}
  {\path{doi:10.1016/0370-2693(92)90491-L}}.

\bibitem{Barr:1992qq}
Stephen~M. Barr and D.~Seckel.
\newblock {Planck scale corrections to axion models}.
\newblock {\em Phys. Rev. D}, 46:539--549, 1992.
\newblock \href {https://doi.org/10.1103/PhysRevD.46.539}
  {\path{doi:10.1103/PhysRevD.46.539}}.

\bibitem{Holdom:1982ex}
Bob Holdom and Michael~E. Peskin.
\newblock {Raising the Axion Mass}.
\newblock {\em Nucl. Phys. B}, 208:397--412, 1982.
\newblock \href {https://doi.org/10.1016/0550-3213(82)90228-0}
  {\path{doi:10.1016/0550-3213(82)90228-0}}.

\bibitem{Dine:1986bg}
Michael Dine and Nathan Seiberg.
\newblock {String Theory and the Strong {CP} Problem}.
\newblock {\em Nucl. Phys. B}, 273:109--124, 1986.
\newblock \href {https://doi.org/10.1016/0550-3213(86)90043-X}
  {\path{doi:10.1016/0550-3213(86)90043-X}}.

\bibitem{Kitano:2021fdl}
Ryuichiro Kitano and Wen Yin.
\newblock {Strong CP problem and axion dark matter with small instantons}.
\newblock {\em JHEP}, 07:078, 2021.
\newblock \href {http://arxiv.org/abs/2103.08598} {\path{arXiv:2103.08598}},
  \href {https://doi.org/10.1007/JHEP07(2021)078}
  {\path{doi:10.1007/JHEP07(2021)078}}.

\bibitem{Agrawal:2017ksf}
Prateek Agrawal and Kiel Howe.
\newblock {Factoring the Strong CP Problem}.
\newblock {\em JHEP}, 12:029, 2018.
\newblock \href {http://arxiv.org/abs/1710.04213} {\path{arXiv:1710.04213}},
  \href {https://doi.org/10.1007/JHEP12(2018)029}
  {\path{doi:10.1007/JHEP12(2018)029}}.

\bibitem{Vafa:1983tf}
C.~Vafa and Edward Witten.
\newblock {Restrictions on Symmetry Breaking in Vector-Like Gauge Theories}.
\newblock {\em Nucl. Phys. B}, 234:173--188, 1984.
\newblock \href {https://doi.org/10.1016/0550-3213(84)90230-X}
  {\path{doi:10.1016/0550-3213(84)90230-X}}.

\bibitem{Vafa:1984xg}
Cumrun Vafa and Edward Witten.
\newblock {Parity Conservation in QCD}.
\newblock {\em Phys. Rev. Lett.}, 53:535, 1984.
\newblock \href {https://doi.org/10.1103/PhysRevLett.53.535}
  {\path{doi:10.1103/PhysRevLett.53.535}}.

\bibitem{Rubakov:1997vp}
V.~A. Rubakov.
\newblock {Grand unification and heavy axion}.
\newblock {\em JETP Lett.}, 65:621--624, 1997.
\newblock \href {http://arxiv.org/abs/hep-ph/9703409}
  {\path{arXiv:hep-ph/9703409}}, \href {https://doi.org/10.1134/1.567390}
  {\path{doi:10.1134/1.567390}}.

\bibitem{Ellis:1978hq}
John~R. Ellis and Mary~K. Gaillard.
\newblock {Strong and Weak CP Violation}.
\newblock {\em Nucl. Phys. B}, 150:141--162, 1979.
\newblock \href {https://doi.org/10.1016/0550-3213(79)90297-9}
  {\path{doi:10.1016/0550-3213(79)90297-9}}.

\bibitem{Khriplovich:1985jr}
I.~B. Khriplovich.
\newblock {Quark Electric Dipole Moment and Induced $\theta$ Term in the
  {Kobayashi-Maskawa} Model}.
\newblock {\em Phys. Lett. B}, 173:193--196, 1986.
\newblock \href {https://doi.org/10.1016/0370-2693(86)90245-5}
  {\path{doi:10.1016/0370-2693(86)90245-5}}.

\bibitem{Khriplovich:1993pf}
I.~B. Khriplovich and A.~I. Vainshtein.
\newblock {Infinite renormalization of Theta term and Jarlskog invariant for CP
  violation}.
\newblock {\em Nucl. Phys. B}, 414:27--32, 1994.
\newblock \href {http://arxiv.org/abs/hep-ph/9308334}
  {\path{arXiv:hep-ph/9308334}}, \href
  {https://doi.org/10.1016/0550-3213(94)90419-7}
  {\path{doi:10.1016/0550-3213(94)90419-7}}.

\bibitem{Berezhiani:2000gh}
Zurab Berezhiani, Leonida Gianfagna, and Maurizio Giannotti.
\newblock {Strong CP problem and mirror world: The Weinberg-Wilczek axion
  revisited}.
\newblock {\em Phys. Lett. B}, 500:286--296, 2001.
\newblock \href {http://arxiv.org/abs/hep-ph/0009290}
  {\path{arXiv:hep-ph/0009290}}, \href
  {https://doi.org/10.1016/S0370-2693(00)01392-7}
  {\path{doi:10.1016/S0370-2693(00)01392-7}}.

\bibitem{Hook:2014cda}
Anson Hook.
\newblock {Anomalous solutions to the strong CP problem}.
\newblock {\em Phys. Rev. Lett.}, 114(14):141801, 2015.
\newblock \href {http://arxiv.org/abs/1411.3325} {\path{arXiv:1411.3325}},
  \href {https://doi.org/10.1103/PhysRevLett.114.141801}
  {\path{doi:10.1103/PhysRevLett.114.141801}}.

\bibitem{Fukuda:2015ana}
Hajime Fukuda, Keisuke Harigaya, Masahiro Ibe, and Tsutomu~T. Yanagida.
\newblock {Model of visible QCD axion}.
\newblock {\em Phys. Rev. D}, 92(1):015021, 2015.
\newblock \href {http://arxiv.org/abs/1504.06084} {\path{arXiv:1504.06084}},
  \href {https://doi.org/10.1103/PhysRevD.92.015021}
  {\path{doi:10.1103/PhysRevD.92.015021}}.

\bibitem{Dimopoulos:2016lvn}
Savas Dimopoulos, Anson Hook, Junwu Huang, and Gustavo Marques-Tavares.
\newblock {A collider observable QCD axion}.
\newblock {\em JHEP}, 11:052, 2016.
\newblock \href {http://arxiv.org/abs/1606.03097} {\path{arXiv:1606.03097}},
  \href {https://doi.org/10.1007/JHEP11(2016)052}
  {\path{doi:10.1007/JHEP11(2016)052}}.

\bibitem{Hook:2019qoh}
Anson Hook, Soubhik Kumar, Zhen Liu, and Raman Sundrum.
\newblock {High Quality QCD Axion and the LHC}.
\newblock {\em Phys. Rev. Lett.}, 124(22):221801, 2020.
\newblock \href {http://arxiv.org/abs/1911.12364} {\path{arXiv:1911.12364}},
  \href {https://doi.org/10.1103/PhysRevLett.124.221801}
  {\path{doi:10.1103/PhysRevLett.124.221801}}.

\bibitem{Dimopoulos:1979pp}
Savas Dimopoulos.
\newblock {A Solution of the Strong {CP} Problem in Models With Scalars}.
\newblock {\em Phys. Lett. B}, 84:435--439, 1979.
\newblock \href {https://doi.org/10.1016/0370-2693(79)91233-4}
  {\path{doi:10.1016/0370-2693(79)91233-4}}.

\bibitem{Gherghetta:2016fhp}
Tony Gherghetta, Natsumi Nagata, and Mikhail Shifman.
\newblock {A Visible QCD Axion from an Enlarged Color Group}.
\newblock {\em Phys. Rev. D}, 93(11):115010, 2016.
\newblock \href {http://arxiv.org/abs/1604.01127} {\path{arXiv:1604.01127}},
  \href {https://doi.org/10.1103/PhysRevD.93.115010}
  {\path{doi:10.1103/PhysRevD.93.115010}}.

\bibitem{Gaillard:2018xgk}
M.~K. Gaillard, M.~B. Gavela, R.~Houtz, P.~Quilez, and R.~Del~Rey.
\newblock {Color unified dynamical axion}.
\newblock {\em Eur. Phys. J. C}, 78(11):972, 2018.
\newblock \href {http://arxiv.org/abs/1805.06465} {\path{arXiv:1805.06465}},
  \href {https://doi.org/10.1140/epjc/s10052-018-6396-6}
  {\path{doi:10.1140/epjc/s10052-018-6396-6}}.

\bibitem{Witten:1982fp}
Edward Witten.
\newblock {An SU(2) Anomaly}.
\newblock {\em Phys. Lett. B}, 117:324--328, 1982.
\newblock \href {https://doi.org/10.1016/0370-2693(82)90728-6}
  {\path{doi:10.1016/0370-2693(82)90728-6}}.

\bibitem{Peskin:1980gc}
Michael~E. Peskin.
\newblock {The Alignment of the Vacuum in Theories of Technicolor}.
\newblock {\em Nucl. Phys. B}, 175:197--233, 1980.
\newblock \href {https://doi.org/10.1016/0550-3213(80)90051-6}
  {\path{doi:10.1016/0550-3213(80)90051-6}}.

\bibitem{Preskill:1980mz}
John Preskill.
\newblock {Subgroup Alignment in Hypercolor Theories}.
\newblock {\em Nucl. Phys. B}, 177:21--59, 1981.
\newblock \href {https://doi.org/10.1016/0550-3213(81)90265-0}
  {\path{doi:10.1016/0550-3213(81)90265-0}}.

\bibitem{Kosower:1984aw}
D.~A. Kosower.
\newblock {SYMMETRY BREAKING PATTERNS IN PSEUDOREAL AND REAL GAUGE THEORIES}.
\newblock {\em Phys. Lett. B}, 144:215--216, 1984.
\newblock \href {https://doi.org/10.1016/0370-2693(84)91806-9}
  {\path{doi:10.1016/0370-2693(84)91806-9}}.

\bibitem{Crewther:1979pi}
R.~J. Crewther, P.~Di~Vecchia, G.~Veneziano, and Edward Witten.
\newblock {Chiral Estimate of the Electric Dipole Moment of the Neutron in
  Quantum Chromodynamics}.
\newblock {\em Phys. Lett. B}, 88:123, 1979.
\newblock [Erratum: Phys.Lett.B 91, 487 (1980)].
\newblock \href {https://doi.org/10.1016/0370-2693(79)90128-X}
  {\path{doi:10.1016/0370-2693(79)90128-X}}.

\bibitem{AxionLimits}
Ciaran O'Hare.
\newblock cajohare/axionlimits: Axionlimits.
\newblock \url{https://cajohare.github.io/AxionLimits/}, July 2020.
\newblock \href {https://doi.org/10.5281/zenodo.3932430}
  {\path{doi:10.5281/zenodo.3932430}}.

\bibitem{Dunsky:2022uoq}
David~I. Dunsky, Lawrence~J. Hall, and Keisuke Harigaya.
\newblock {Dark Radiation Constraints on Heavy QCD Axions}.
\newblock 5 2022.
\newblock \href {http://arxiv.org/abs/2205.11540} {\path{arXiv:2205.11540}}.

\bibitem{deSalas:2015glj}
P.~F. de~Salas, M.~Lattanzi, G.~Mangano, G.~Miele, S.~Pastor, and O.~Pisanti.
\newblock {Bounds on very low reheating scenarios after Planck}.
\newblock {\em Phys. Rev. D}, 92(12):123534, 2015.
\newblock \href {http://arxiv.org/abs/1511.00672} {\path{arXiv:1511.00672}},
  \href {https://doi.org/10.1103/PhysRevD.92.123534}
  {\path{doi:10.1103/PhysRevD.92.123534}}.

\bibitem{Aloni:2018vki}
Daniel Aloni, Yotam Soreq, and Mike Williams.
\newblock {Coupling QCD-Scale Axionlike Particles to Gluons}.
\newblock {\em Phys. Rev. Lett.}, 123(3):031803, 2019.
\newblock \href {http://arxiv.org/abs/1811.03474} {\path{arXiv:1811.03474}},
  \href {https://doi.org/10.1103/PhysRevLett.123.031803}
  {\path{doi:10.1103/PhysRevLett.123.031803}}.

\bibitem{Goudzovski:2022vbt}
Evgueni Goudzovski et~al.
\newblock {New Physics Searches at Kaon and Hyperon Factories}.
\newblock 1 2022.
\newblock \href {http://arxiv.org/abs/2201.07805} {\path{arXiv:2201.07805}}.

\bibitem{Kelly:2020dda}
Kevin~J. Kelly, Soubhik Kumar, and Zhen Liu.
\newblock {Heavy axion opportunities at the DUNE near detector}.
\newblock {\em Phys. Rev. D}, 103(9):095002, 2021.
\newblock \href {http://arxiv.org/abs/2011.05995} {\path{arXiv:2011.05995}},
  \href {https://doi.org/10.1103/PhysRevD.103.095002}
  {\path{doi:10.1103/PhysRevD.103.095002}}.

\bibitem{Ertas:2020xcc}
Fatih Ertas and Felix Kahlhoefer.
\newblock {On the interplay between astrophysical and laboratory probes of
  MeV-scale axion-like particles}.
\newblock {\em JHEP}, 07:050, 2020.
\newblock \href {http://arxiv.org/abs/2004.01193} {\path{arXiv:2004.01193}},
  \href {https://doi.org/10.1007/JHEP07(2020)050}
  {\path{doi:10.1007/JHEP07(2020)050}}.

\bibitem{Chakraborty:2021wda}
Sabyasachi Chakraborty, Manfred Kraus, Vazha Loladze, Takemichi Okui, and
  Kohsaku Tobioka.
\newblock {Heavy QCD axion in b\textrightarrow{}s transition: Enhanced limits
  and projections}.
\newblock {\em Phys. Rev. D}, 104(5):055036, 2021.
\newblock \href {http://arxiv.org/abs/2102.04474} {\path{arXiv:2102.04474}},
  \href {https://doi.org/10.1103/PhysRevD.104.055036}
  {\path{doi:10.1103/PhysRevD.104.055036}}.

\bibitem{Bertholet:2021hjl}
Emilie Bertholet, Sabyasachi Chakraborty, Vazha Loladze, Takemichi Okui, Abner
  Soffer, and Kohsaku Tobioka.
\newblock {Heavy QCD axion at Belle II: Displaced and prompt signals}.
\newblock {\em Phys. Rev. D}, 105(7):L071701, 2022.
\newblock \href {http://arxiv.org/abs/2108.10331} {\path{arXiv:2108.10331}},
  \href {https://doi.org/10.1103/PhysRevD.105.L071701}
  {\path{doi:10.1103/PhysRevD.105.L071701}}.

\bibitem{Chou:2016lxi}
John~Paul Chou, David Curtin, and H.~J. Lubatti.
\newblock {New Detectors to Explore the Lifetime Frontier}.
\newblock {\em Phys. Lett. B}, 767:29--36, 2017.
\newblock \href {http://arxiv.org/abs/1606.06298} {\path{arXiv:1606.06298}},
  \href {https://doi.org/10.1016/j.physletb.2017.01.043}
  {\path{doi:10.1016/j.physletb.2017.01.043}}.

\bibitem{Craig:2018kne}
Nathaniel Craig, Anson Hook, and Skyler Kasko.
\newblock {The Photophobic ALP}.
\newblock {\em JHEP}, 09:028, 2018.
\newblock \href {http://arxiv.org/abs/1805.06538} {\path{arXiv:1805.06538}},
  \href {https://doi.org/10.1007/JHEP09(2018)028}
  {\path{doi:10.1007/JHEP09(2018)028}}.

\bibitem{Bedi:2022qrd}
Ravneet~S. Bedi, Tony Gherghetta, and Maxim Pospelov.
\newblock {Enhanced EDMs from Small Instantons}.
\newblock 5 2022.
\newblock \href {http://arxiv.org/abs/2205.07948} {\path{arXiv:2205.07948}}.

\end{thebibliography}

\bibliographystyle{unsrturl}

\end{document}